\pdfoutput=1
\documentclass[12pt]{article}
\usepackage{amsmath}
\usepackage{amsfonts}
\usepackage{amssymb}
\usepackage{graphicx}
\graphicspath{{figures/}}
\usepackage{multirow}
\usepackage{array}
\usepackage{xcolor}
\usepackage{algorithm}
\usepackage{algpseudocode}
\usepackage[authoryear,round]{natbib}
\usepackage{hyperref}
\hypersetup{hidelinks}
\usepackage{bigints}
\usepackage{relsize,exscale}
\usepackage{subcaption}  
\usepackage{bm}
\usepackage{multirow}
\usepackage{booktabs}
\usepackage{tabularx}
\usepackage{booktabs}
\usepackage{arydshln}
\usepackage{float} 
\usepackage{url}
\usepackage{pdfpages}
\usepackage{pdflscape}
\usepackage{dcolumn}
\usepackage{caption}
\usepackage{makecell}
\usepackage{float}
\usepackage{siunitx}
\usepackage{fancybox}
\usepackage{appendix}
\usepackage{chngcntr}
\usepackage{orcidlink}
\usepackage{fancyhdr}
\usepackage[a4paper, total={6in, 8.5in}]{geometry} 
\title{Exact association-test distributions for prior-informed stepwise SNP prioritization}

\author{N. Kossi Doulabe\orcidlink{0009-0005-6317-5654}\textsuperscript{1}, Anne-Sophie Charest\textsuperscript{1,2}, Lajmi Lakhal-Chaieb\orcidlink{0000-0002-8358-4392}\textsuperscript{1,*} \\
\small \textsuperscript{1}Département de Mathématiques et de Statistique, Université Laval, Québec, Canada \\
\small \textsuperscript{2} Institut Intelligence et Données, Université Laval, Québec, Canada \\
    \small \textsuperscript{*}Correspondence: Lajmi Lakhal-Chaieb (Lajmi.Lakhal@mat.ulaval.ca)} 
\date{}
\begin{document}

\maketitle
\begin{abstract}
\noindent Prioritizing causal variants in Genome-Wide Association Studies (GWAS) is challenging due to linkage disequilibrium and high dimensionality. We propose an iterative algorithm for the selection of genetic variables that allows the integration of biological priors. The proposed algorithm requires the specification of the power function of a test statistic and incorporates {\it a priori} information on each SNP. We derive the approximate and exact distributions of several commonly used test statistics. At each step of the proposed algorithm, we select the SNP with the largest {\it a posteriori} probability of being causal. We consider several stopping criteria for the algorithm. We conduct comprehensive simulation studies to assess the performance of several versions of the proposed algorithm and to make a comparison to some existing competitors. The use of the proposed prioritization algorithm is illustrated with a cis-eQTL application to a human gene-expression data, using European discovery and Yoruba cross-population replication.
\end{abstract}


\newpage
\section{Introduction}

\noindent Genome-wide association studies (GWAS) have identified thousands of loci associated with complex traits and diseases, but the lead SNP at an associated locus is rarely guaranteed to be causal. Linkage disequilibrium (LD) distributes the association signal across correlated markers, so a detected SNP may be a tag variant, a proxy for an untyped causal allele, or a false positive \citep{gogele2012,hormozdiari2014,hormozdiari2015,schaid2018,hautakangas2024,wu2024,long2025}. This creates a methodological gap between association discovery and biological interpretation. GWAS points to regions, whereas downstream experimental and translational work requires a much shorter list of plausible causal variants.\\

\noindent Statistical fine-mapping addresses this gap by modeling multiple candidate variants and reporting variant-level measures of support, usually through Bayesian posterior inclusion probabilities (PIPs) or credible sets \citep{guan2011,wen2014,stephens2009,the_wellcome_trust_case_control_consortium_bayesian_2012,wang2020,li2024}. Modern fine-mapping tools, including CAVIAR \citep{hormozdiari2014}, FINEMAP \citep{benner2016}, SuSiE \citep{wang2020} and CARMA \citep{yang2023}, have greatly improved locus-level resolution and scalability. Other Bayesian regression and shrinkage approaches, such as varbvs \citep{carbonetto2012}, provide complementary strategies for high-dimensional genotype data \citep{Logsdon2010,Habier2011,zhou2013,ge2019,song2022}. Functional annotations can further improve prioritization when they are relevant to the trait and tissue, as shown in annotation-informed fine-mapping and post-GWAS integration studies \citep{kichaev2014,pickrell2014,roadmap2015,chen2016,karimnezhad2020,he2022}.\\

\noindent Prior-informed ranking of candidate variants arises in at least three practical settings. First, follow-up resources are limited. A study carries a fixed budget of $K$ variants forward to functional validation or targeted genotyping, so the operational question within an associated region is which variants to pursue first. Ordering hypotheses by external information is the rationale of weighted hypothesis testing \citep{genovese2006,roeder2009} and of functionally informed re-weighting such as FINDOR \citep{kichaev2019}. Second, conditional stepwise scans are standard practice for separating signals within an associated region, with each candidate scored given the currently selected set \citep{yang2012}. The framework developed here equips this conditional path with finite-sample calibration and explicit prior weights. Finally, combining association evidence with functional annotation is central to fine-mapping follow-up, including expression quantitative trait locus (eQTL) studies in which each gene defines a natural candidate window. These settings share the same underlying problem, namely ranking candidates under a limited follow-up budget while incorporating potentially imperfect prior information, often at sample sizes for which asymptotic calibration may be unreliable.\\

\noindent Two statistical issues remain especially important for stepwise or screening-oriented procedures. First, many association statistics are approximated by standard normal or chi-square distributions, implicitly ignoring the variability of the estimated residual variance. This is convenient, but it can miscalibrate evidence when the model is refit many times along an expanding selected set \citep{zeng2017,banerjee2018,zhu2017}. Second, biological priors are not uniformly reliable. A high-quality annotation sharpens the ranking, but a misspecified prior actively pushes a method toward the wrong part of the genome, acutely so for joint selection rules that amplify prior information rather than merely using it as a weak tie-breaker.\\

\noindent We propose a prior-informed stepwise SNP-prioritization framework that directly targets these two issues. At each step, the method evaluates candidate SNPs through a local posterior score obtained by combining a SNP-level annotation weight with the finite-sample distribution of a Wald, Score or joint multiple-SNP statistic. The score is path-conditional and local. It is a greedy evidence score designed for computationally light prioritization. Exact non-central $t$ and $F$ distributions are used when available, and are compared with naive and first-order asymptotic approximations.\\

\noindent This framework sits at the intersection of several literatures, but its objective differs from each. Annotation-informed fine-mapping targets model-averaged inclusion probabilities within a locus, whereas our procedure produces path-conditional scores along a single greedy path. Classical forward selection carries no prior weight and relies on nominal normal or chi-square distributions. Bayesian variable selection and penalized regression fit a full joint model over the variants. Selective-inference methods target valid post-selection intervals rather than ranking. And weighted multiple-testing procedures re-weight marginal $p$-values once \citep{genovese2006,kichaev2019}, whereas our scores are recomputed conditionally along the path, so the prior informs every conditional decision rather than only the initial screen. The method thus complements locus-level fine-mapping tools and full Bayesian model averaging as a prior-sensitive stepwise prioritization procedure.\\

\noindent The paper makes five contributions, the first two of which are primarily theoretical. First, we derive the exact and approximate finite-sample distributions of the single-SNP Wald and Score statistics, and of the joint Multiple-SNP statistic, in the path-conditional stepwise setting in which each candidate is scored given the currently selected set. These distributions explicitly account for the variability of the residual-variance estimate that the usual normal and chi-square approximations treat as fixed, and are summarized in Table~\ref{tab:distributions}. Second, we show how these distributions embed into a prior-weighted local posterior score for greedy prioritization, whose prior integrals are computed exactly. Because they calibrate candidate statistics conditional on a fixed selected set, the same distributions can be reused outside this Bayesian score, for instance in classical forward selection or conditional screening. Third, we quantify the practical payoff of this finite-sample calibration. It is substantial for the single-variant Wald statistic at small sample sizes, where the normal approximations are weak, and negligible for the joint statistic. Fourth, we show through the broader simulations that, once the sample is large, prior quality rather than the choice of statistic becomes the main determinant of joint-method performance. A faithful prior is decisive, and a misspecified prior is actively harmful. Fifth, we compare fixed-$K$ and adaptive stopping rules (adjusted $R^2$, BIC, eBIC and posterior-score thresholds) and assess the framework against some standard methods. The methodology is evaluated at three levels. The first is a set of simulations based on real and on parametrically controlled LD structures, including the small-sample regime where the exact distributions depart from their asymptotic approximations. The second is a semi-synthetic evaluation built on real human genotypes from the Genetic European Variation in Health and Disease (GEUVADIS) study \citep{lappalainen2013,the1000genomes2015}, with known simulated truths and PolyFun-derived functional prior weights \citep{weissbrod2020}. The third is a real cis-eQTL fine-mapping application to GEUVADIS gene expression, with priors spanning a gradient of biological relevance, from discordant to proximity to the transcription start site (TSS), and a cross-population replication.\\

\noindent The remainder of the paper presents the stepwise algorithm, the test-statistic distributions and the stopping rules in Section 2, the simulation evidence, with complete numerical grids in the Supplement (Section 3), the GEUVADIS cis-eQTL application in Section 4, and practical use, limitations and extensions in Section 5.

\section{Materials and methods}

\subsection{Data structure and variables}
\noindent In genome-wide association studies (GWAS), millions of genetic variants are measured for a large number of individuals to detect genomic regions associated with a human characteristic or disease. Single nucleotide polymorphisms (SNPs) are probably the most common type of genetic variant. In our study, we consider $n$ independent individuals and for individual $i$, $i=1,\cdots,n$, we observe a set of $m$ SNPs $\{G_{i1},\cdots,G_{im}\}$ and a quantitative phenotype $Y_i$. We assume that $n \ll m$, i.e., the number of SNPs is considerably larger than the sample size. Furthermore, we assume that there exist $m$ latent standardized effect sizes $\lambda_j$, $j=1,\ldots,m$, such that $\lambda_j \neq 0$ indicates that SNP $j$ contributes to the phenotype, whereas $\lambda_j = 0$ corresponds to a null effect. Following the Bayesian variable-selection literature \citep{Logsdon2010,guan2011,Habier2011}, we assume a spike-and-slab prior
$$\lambda_j \sim (1-p_j)\,\delta_0 + p_j\,\mathcal{N}(0,\tau_j^2),$$
\noindent where $\delta_0$ denotes a point mass at zero, $p_j \in [0,1]$ is the prior inclusion probability for SNP $j$, and $\tau_j^2>0$ controls the prior variance of non-null effects. In other words,
$$P(\lambda_j \neq 0) = p_j, \qquad P(\lambda_j = 0) = 1-p_j, \mbox{ and } \qquad \lambda_j|\lambda_j \neq 0 \sim {\cal N}(0,\tau_j^2).$$
\noindent In practice, $p_j$ may incorporate biological annotation information, previous GWAS evidence, functional scores, or expert knowledge. The hyperparameters $\{(p_j,\tau_j^2)\}_{j=1,\ldots,m}$ are assumed to be known. Within the stepwise procedure developed below, $p_j$ is a \emph{local annotation weight} entering the candidate-level Bayes score at each step, rather than a genome-wide sparse prior. A value such as $p_j = 0.5$ or $0.7$ encodes the relative annotation strength of SNP $j$ at the point where it is scored, and does not imply the $m\,\bar p$ expected causal variants that the same value would carry as a global marginal inclusion probability across all $m$ SNPs.
\noindent The SNPs data $G_{ij} \in \{0,1,2\}$, $i=1,\cdots,n$, $j=1,\cdots,m$, are counts of minor alleles for SNP $j$ and individual $i$. For ease of computation and without loss of generality, we scale these counts into 
$$X_{ij}= \frac{G_{ij} - \bar{G}_j}{\sqrt{S_{G_j}^2}},$$
where
$$ \bar{G}_j = \frac{1}{n} \sum_{i=1}^n  G_{ij}$$
and 
$$S_{G_j}^2 = \frac{1}{n} \sum_{i=1}^n  (G_{ij} - \bar{G}_j)^2.$$
Thus, we have $\displaystyle \sum_{i=1}^n X_{ij} = 0$ and $\displaystyle\sum_{i=1}^n X_{ij}^2 = n$.

\noindent Therefore, the data at hand is $(Y,X)$, where $Y=(Y_1,\cdots,Y_n)^\top$ is the vector of phenotypes, $A^\top$ denotes the transpose of $A$ and $X$ is a $n \times m$ matrix of scaled SNPs. Given the data and hyperparameters $\{(p_j,\tau^2_j),j=1,\cdots,m\}$, we aim to select a set of causal SNPs that are associated with the phenotype. To this end, we develop an iterative algorithm and present, assess and compare several variants of the proposed procedure. The main contribution of this work is to propose, implement, assess, and apply an iterative algorithm dedicated to this task.

\subsection{SNP selection procedure}

The selection procedure consists of successive steps. At each step, we select the highest-scoring candidate SNP. The algorithm stops after $K$ steps, where $K$ can be a predetermined number or selected by a stopping criterion. For $k=0,1,\cdots,K$, we denote by $S_k$ the set of the indices of the SNPs already selected at the end of step $k$ and we let $T_k=\{1,\cdots,m\} \setminus S_k$. Note that $T_k$ is composed of the indices of the SNPs still to be examined after step $k$. Therefore, one has $S_0=\emptyset$, $T_0=\{1,\cdots,m\}$, $\mbox{Card}(S_k)=k$, and $\mbox{Card}(T_k)=m-k$. The outcome of the iterative algorithm is $S_K$.

\subsubsection{Iterative algorithm}

At step $k=1$, for $j \in T_0$, we consider the following linear model
\begin{equation}\label{model1}
    Y = a_n \beta_{0j1} + X_{*j}\beta_{1j1} + \varepsilon_{j1},
\end{equation}
where $a_n = (1, 1, \ldots, 1)^{\top}$, $X_{*j}$ is the $j^{\mbox{th}}$ column of $X$, $\beta_{0j1}$ the intercept, $\beta_{1j1}$ the regression coefficient for SNP $j$, $\varepsilon_{j1} = (\varepsilon_{1j1}, \varepsilon_{2j1}, \ldots, \varepsilon_{nj1})^\top \sim \mathcal{N}(0, \sigma^2_{j1}I_n)$ the vector of error terms and $I_n$ the $n$ by $n$ identity matrix. We are interested in testing the null hypothesis $\mathbb{H}_0^{(j1)}: \beta_{1j1} = 0$.

\noindent Let $Z_{j1}$ be a test statistic for $\mathbb{H}_0^{(j1)}$ and denote by $h_{j1}(\cdot;\lambda_j)$ its density function given the effect size $\lambda_j$. In the next section, we consider several test statistics, and for each statistic, we show that $h_{j1}$ depends only on the scaled regression coefficient $\beta_{1j1}/\sigma_{j1}$. This prompts us to specify the prior distribution for this parameter and set $\lambda_j=\beta_{1j1}/\sigma_{j1}$. The null hypothesis can therefore be written as $\mathbb{H}_0^{(j1)}: \lambda_j = 0$. Given the observed value of the test statistic $Z_{j1}=z_{j1}$, we score SNP $j$ by its stepwise posterior score $q_{j1}$, a local measure of inclusion evidence conditional on the current selected path. This score is distinct from the posterior inclusion probability (PIP) of ensemble fine-mappers such as SuSiE and FINEMAP, which is a model-averaged inclusion probability over the full $2^m$ model space. Instead, $q_{j1}$ is a sequential, path-conditional quantity, and we keep the two terms separate throughout:
\begin{equation}\label{postproba}
q_{j1} = P(\lambda_j \neq 0 \mid Z_{j1}=z_{j1}) = \frac{p_j\,A_{j1}}{p_j\,A_{j1}+(1-p_j)\,B_{j1}},
\end{equation}
where
\begin{eqnarray*}
A_{j1} &=& \int_{-\infty}^{\infty} h_{j1}(z_{j1};\lambda_j)\,\frac{1}{\tau_j}\, \phi\!\left(\frac{\lambda_j}{\tau_j}\right) d\lambda_j, \\
B_{j1} &=& h_{j1}(z_{j1};\lambda_j=0),
\end{eqnarray*}
and $\phi(\cdot)$ is the density function of the standard normal distribution. Note that $A_{j1}$ is the marginal density of $Z_{j1}$ under the non-null component evaluated at $z_{j1}$, whereas $B_{j1}$ is the density under the null. We compute $\{q_{j1}, j \in T_0\}$ and denote by $l_1$ the index of the SNP with the largest stepwise posterior score. Therefore, one has $S_1=\{l_1\}$ and $T_1 = \{1,\cdots,m\}\setminus S_1 = \{1,2,\cdots,l_1-1,l_1+1,\cdots,m\}$.

\noindent At the beginning of step $k, k\ge 2$, for each $j \in T_{k-1}$, we consider the following linear model

\begin{equation}\label{modelk}
    Y = a_n \beta_{0jk} + X_{*j}\beta_{1jk} + X^{(k)} \beta_{2jk} + \varepsilon_{jk},
\end{equation}
where $\beta_{0jk}$, $\beta_{1jk}$ and $\beta_{2jk}$ are regression coefficients, $X^{(k)}=\{X_{il}\}_{1 \le i \le n; l \in S_{k-1}}$ is the genetic data of the already selected SNPs ($k-1$ columns, so that model~\eqref{modelk} contains $k$ genetic predictors and $n-k-1$ residual degrees of freedom) and $\varepsilon_{jk} = (\varepsilon_{1jk}, \varepsilon_{2jk}, \ldots, \varepsilon_{njk})^{\top} \sim \mathcal{N}(0, \sigma^2_{jk}I_n)$ is a vector of error terms. In the above equation, $\beta_{0jk}$ and $\beta_{1jk}$ are scalar parameters whereas $\beta_{2jk}$ is a vector of length $k-1$. We are interested in testing $\mathbb{H}_0^{(jk)}: \beta_{1jk} = 0.$ \\

\noindent Let $Z_{jk}$ be a test statistic for $\mathbb{H}_0^{(jk)}$ and denote its density function by $h_{jk}$. In the next section, we show that $h_{jk}$ can be expressed as $h_{jk}(\cdot;\lambda_{jk},W_k,r_{jk},\rho_{jk})$, where $\lambda_{jk}=\beta_{1jk}/\sigma_{jk}$ is the standardized partial effect of SNP $j$ given the already selected SNPs (at step $k=1$, $\lambda_{j1}=\lambda_j$), $W_k=n^{-1}{X^{(k)}}^\top X^{(k)}$ and $r_{jk} = n^{-1}{X^{(k)}}^\top X_{*j}$ are empirical correlations between the SNPs considered in the above model, and $\rho_{jk}=r_{jk}^\top W_k^{-1} r_{jk} \in [0,1]$ is the squared multiple correlation coefficient, that is, the coefficient of determination of the regression of $X_{*j}$ on the columns of $X^{(k)}$. It assesses how well $X_{*j}$ can be predicted as a linear combination of these columns. The partial effect $\lambda_{jk}$ depends on the conditioning set $S_{k-1}$ and the stepwise score applies the spike-and-slab prior introduced above to $\lambda_{jk}$, with the same hyperparameters $(p_j,\tau_j^2)$ at every step.

\noindent For each $j \in T_{k-1}$, we define the stepwise posterior score as
$$q_{jk}=P(\lambda_{jk} \neq 0 \mid Z_{jk}=z_{jk}) = \frac{p_j\,A_{jk}}{p_j\,A_{jk}+(1-p_j)\,B_{jk}},$$
where
\begin{eqnarray*}
A_{jk} &=& \int_{-\infty}^{\infty} h_{jk}(z_{jk};\lambda_{jk},W_k,r_{jk},\rho_{jk})\,\frac{1}{\tau_j}\, \phi\!\left(\frac{\lambda_{jk}}{\tau_j}\right) d\lambda_{jk}, \\
B_{jk} &=& h_{jk}(z_{jk};\lambda_{jk}=0,W_k,r_{jk},\rho_{jk}).
\end{eqnarray*}
We denote by $l_k$ the index of the SNP with the largest $q_{jk}$. At the end of step $k$, we have $S_k=\{l_1,\cdots,l_k\}=S_{k-1} \cup \{l_k\}$ and $T_k = T_{k-1} \setminus \{l_k\}$.

\noindent The algorithm stops after $K$ steps and the output is $S_K$, which is the set of indices of the selected SNPs.

\noindent The procedure is a forward greedy scan driven by this prior-weighted score. Algorithm~\ref{alg:stepwise} states the loop in pseudocode.

\begin{algorithm}[htbp]
\caption{Prior-informed stepwise SNP prioritization}\label{alg:stepwise}
\begin{algorithmic}[1]
\State \textbf{Input:} genotypes $X$, phenotype $Y$; priors $\{(p_j,\tau_j)\}_{j=1}^m$; statistic family (Wald / Score / Multiple; naive, asymptotic or exact); fixed $K$ or stopping rule
\State Initialize selected set $S_0 \gets \emptyset$, candidate set $T_0 \gets \{1,\dots,m\}$, step $k \gets 1$
\Repeat
  \For{each candidate $j \in T_{k-1}$}
    \State Residualize on $S_{k-1}$: form $W_k,\ r_{jk},\ \rho_{jk}$ and the conditional statistic $Z_{jk}$
    \State Evaluate its finite-sample density $h_{jk}$ (Table~\ref{tab:distributions})
    \State $q_{jk} \gets$ combine $h_{jk}$ with the prior weight $p_j$, integrated over the slab scale $\tau_j$
  \EndFor
  \State $l_k \gets \arg\max_{j \in T_{k-1}} q_{jk}$ \Comment{highest-scoring candidate}
  \State $S_k \gets S_{k-1}\cup\{l_k\}$;\quad $T_k \gets T_{k-1}\setminus\{l_k\}$;\quad $k \gets k+1$
\Until{fixed $K$ reached \textbf{or} stopping rule met (Adj.\ $R^2$ / BIC / eBIC / posterior-score threshold)}
\State \Return selected set $S_k$ and its stepwise scores $\{q_{l_1},\dots,q_{l_k}\}$
\end{algorithmic}
\end{algorithm}

\subsubsection{Test statistics and their distributions}\label{sect2.2.2}

In this section, we consider the Wald and Score statistics to assess the significance of the null hypothesis $\mathbb{H}_0^{(jk)}$ and consider different approaches to derive their distributions.

\noindent In Supplementary Appendix $A_1$, we show that, for $k=1$, the Wald and the Score statistics are given by 

$$Z_{j1}^{(W)} = \frac{X_{*j}^\top Y}{\sqrt{n\hat\sigma^2_{j1}}}$$
and 

$$Z_{j1}^{(S)} = \frac{X_{*j}^\top Y}{\sqrt{n\tilde\sigma^2_{j1}}},$$
where
$$\hat\sigma^2_{j1} = \frac{1}{n-2} Y^\top\left(I_n-\frac{1}{n} a_na_n^\top - \frac{1}{n}  X_{*j} X_{*j}^\top\right) Y$$
and 
$$\tilde\sigma^2_{j1} = \frac{1}{n-1} Y^\top\left(I_n-\frac{1}{n} a_na_n^\top \right) Y$$
are the REML estimates of $\sigma^2_{j1}$, in the general case and under $\mathbb{H}_0^{(j1)}$, respectively. 

\noindent A common practice among scientists is to ignore the variability of the REML estimate of $\sigma^2_{j1}$ and approximate the distribution of $Z_{j1}^{(W)}$ and $Z_{j1}^{(S)}$ by ${\cal N}(\sqrt{n}\lambda_j,1)$ \citep{hormozdiari2015}. We refer to this strategy as the naive Wald (NaW) and Score (NaS) approaches in the sequel.

\noindent In Supplementary Appendix $A_2$, we show that the asymptotic distributions of $Z_{j1}^{(W)}$ and $Z_{j1}^{(S)}$ are respectively, given by
$$Z_{j1}^{(W)} \sim {\cal N}\left(\sqrt{n}\lambda_j\, , \,1+\frac{1}{2} \lambda_j^2\right)$$
and
$$Z_{j1}^{(S)} \sim {\cal N}\left\{\frac{\sqrt{n}\lambda_j}{\sqrt{1+\lambda_j^2}}\, , \,\frac{1}{(1+\lambda_j^2)^3}\left(1+\frac{1}{2}\lambda_j^2\right)\right\}.$$
We refer to these results as asymptotic Wald (AsW) and Score (AsS) approaches in the sequel.\\

\noindent Following \citet[section~5.4.2]{Chapter5Rencher}, we show that the exact distribution of $Z_{j1}^{(W)}$ is a non-central $t$ distribution with $n-2$ degrees of freedom and a parameter of non-centrality equal to $\sqrt{n}\lambda_j$, i.e., $Z_{j1}^{(W)} \sim t_{n-2}(\sqrt{n}\lambda_j)$. This result is referred to as the exact Wald (ExW) approach. Throughout, the qualifier \emph{exact} denotes a distribution that holds exactly in finite samples under the Gaussian homoscedastic linear model~(\ref{model1}), with independent errors and under the null as well as under the alternative, as opposed to the large-sample normal approximation used by the asymptotic and naive variants. It is not a claim of robustness against violations of these assumptions (heavy-tailed phenotypes, residual population structure), which must be addressed by the usual modeling and diagnostic steps.

\noindent This exact distribution follows directly from the sampling structure of the marginal fit, in which $\hat\beta_{1j1}\sim\mathcal N(\beta_{1j1},\,\sigma^2_{j1}/n)$ and $(n-2)\hat\sigma^2_{j1}/\sigma^2_{j1}\sim\chi^2_{n-2}$ are independent, so the studentized statistic $Z_{j1}^{(W)}=\sqrt{n}\,\hat\beta_{1j1}/\hat\sigma_{j1}$ is, by definition, a non-central $t_{n-2}$ with a non-centrality parameter equal to $\sqrt{n}\,\lambda_j$. The naive and asymptotic variants replace the exact $\chi^2$ term by a fixed plug-in or by a first-order moment expansion of $\hat\sigma^2_{j1}$, which recovers the normal distributions stated above. The full algebra is given in Supplementary Appendix~A2.

\noindent Finally, using algebraic computations, we show that
$$\frac{1}{{Z_{j1}^{(W)}}^2} = \frac{n}{n-2} \left(\frac{n-1}{n}\times \frac{1}{{Z_{j1}^{(S)}}^2}-\frac{1}{n}\right)  $$

\noindent Since there is a one-to-one relationship between ${Z_{j1}^{(W)}}^2$ and ${Z_{j1}^{(S)}}^2$, the exact score approach gives posterior probabilities identical to the exact Wald approach. \\

\noindent In the Supplementary Appendix $B_1$, we show that, for $k \ge 2$, the Wald and Score statistics are given, respectively, by
$$Z_{jk}^{(W)} = \frac{(X_{*j}-X^{(k)}W_k^{-1}r_{jk})^\top Y}{\sqrt{n(1-\rho_{jk})\hat\sigma^2_{jk}}}$$
and

$$Z_{jk}^{(S)} = \frac{(X_{*j}-X^{(k)}W_k^{-1}r_{jk})^\top Y}{\sqrt{n(1-\rho_{jk})\tilde\sigma^2_{jk}}}.$$
\noindent Here, $\hat\sigma^2_{jk}$ and $\tilde\sigma^2_{jk}$ are estimates of $\sigma^2_{jk}$, in the general case and under $\mathbb{H}_0^{(jk)}$, respectively. These estimates are equal to
$$\hat\sigma^2_{jk} = \frac{1}{n-(k+1)} Y^\top P_{jk} Y$$
and
$$\tilde\sigma^2_{jk} = \frac{1}{n-k} Y^\top P^0_{jk} Y,$$
where
$$P^0_{jk} = I_n - \frac{1}{n} a_n a_n^\top - \frac{1}{n} X^{(k)} W_k^{-1} X^{(k)\top}$$
and
$$P_{jk} = P^0_{jk}  - \displaystyle\frac{(X^{(k)} W_k^{-1} r_{jk} - X_{*j})(X^{(k)} W_k^{-1} r_{jk} - X_{*j})^{\top}}{n(1 - \rho_{jk})}$$
are the projection matrices.

\noindent If one ignores the variability of $\hat\sigma^2_{jk}$ and $\tilde\sigma^2_{jk}$, it is straightforward to see that the naive distributions of $Z_{jk}^{(W)}$ and $Z_{jk}^{(S)}$ are ${\cal N}\{\sqrt{n(1-\rho_{jk})}\lambda_{jk}\, , \, 1\}$.

\noindent In the Supplementary Appendix $B_2$, we show that $Z_{jk}^{(W)}$ and $Z_{jk}^{(S)}$ converge in distribution to 
$${\cal N}\left\{\sqrt{n(1-\rho_{jk})}\lambda_{jk} \, , \, 1+\frac{(1-\rho_{jk})\lambda^2_{jk}}{2}\right\}$$
and
$${\cal N}\left\{\sqrt{\frac{n(1-\rho_{jk})}{1+(1-\rho_{jk})\lambda_{jk}^2}}\lambda_{jk} \, , \, \frac{2+(1-\rho_{jk})\lambda_{jk}^2}{2\left[1+(1-\rho_{jk})\lambda_{jk}^2\right]^3}\right\},$$
respectively.
\noindent We also show that the exact distribution of $Z_{jk}^{(W)}$ is a non-central $t$ distribution with $n-k-1$ degrees of freedom and a non-centrality parameter equal to $\lambda_{jk}\sqrt{n(1-\rho_{jk})}$. This mirrors the marginal derivation, now with a residualized effect and an LD deflation factor, since $\hat\beta_{1jk}\sim\mathcal N\!\big(\beta_{1jk},\,\sigma^2_{jk}/[n(1-\rho_{jk})]\big)$ and $(n-k-1)\hat\sigma^2_{jk}/\sigma^2_{jk}\sim\chi^2_{n-k-1}$ independently, so $Z_{jk}^{(W)}\sim t_{n-k-1}\!\big(\lambda_{jk}\sqrt{n(1-\rho_{jk})}\big)$. The degrees of freedom drop by one at each step and the factor $1-\rho_{jk}$ discounts the information already carried by the selected set. For the joint Multiple-SNP statistic of the next section, the same argument applied to a $k$-dimensional quadratic form yields a non-central $F_{k,\,n-k-1}$ distribution.

\noindent Finally, we demonstrate that there is a one-to-one relationship between $Z_{jk}^{(W)}$ and $Z_{jk}^{(S)}$. Therefore, the posterior probabilities computed based on the exact distributions of $Z_{jk}^{(W)}$ and $Z_{jk}^{(S)}$ are identical.

\noindent The difference between the naive, asymptotic and exact forms lies entirely in how each treats the estimated residual variance $\widehat\sigma^2_{jk}$. The naive form fixes it at a plug-in value and refers the standardized statistic to a normal distribution with unit variance (or, jointly, to a non-central chi-square). The asymptotic form keeps the first-order normal limit but corrects its variance for the estimation of $\sigma^2_{jk}$. The exact form integrates this estimation uncertainty exactly, which is precisely what turns the normal into a non-central $t$ in the marginal case and the chi-square into a non-central $F$ in the joint case, with degrees of freedom $n-k-1$ that shrink as the selected path grows. Table \ref{tab:distributions} collects the resulting distributions.

\begin{table}[htbp] 
\centering 
\caption{Test-statistic distributions used in the stepwise score.} 
\label{tab:distributions}
\small 
\renewcommand{\arraystretch}{1.35} 
\setlength{\tabcolsep}{6pt} 
\begin{tabular}
{ >{\raggedright\arraybackslash}p{3.0cm} >{\centering\arraybackslash}p{1.5cm} >{\raggedright\arraybackslash}p{6.8cm} >{\centering\arraybackslash}p{2.1cm} }
\hline
\textbf{Statistic} & \textbf{Step} & \centering\textbf{Distribution under $\lambda$} & \textbf{Method} \tabularnewline
\hline 
Marginal Wald & $k=1$ & Non-central \(t_{n-2}\left(\sqrt{n}\lambda_j\right)\) & ExW \tabularnewline Marginal Score & $k=1$ & One-to-one map to the marginal Wald statistic & AsS \tabularnewline Conditional Wald & $k\geq 2$ & Non-central \(t_{n-k-1}\left( \lambda_{jk}\sqrt{n(1-\rho_{jk})} \right)\) & ExW \tabularnewline Conditional Score & $k\geq 2$ & Asymptotic normal with residual-variance correction & AsS \tabularnewline Multiple-SNP & $k\geq 2$ & Non-central \(F_{k,n-k-1}\left(\widetilde{\delta}_{jk}\right)\) (exact prior-integrated score) & ExM~/~NaM \tabularnewline \hline \end{tabular} \end{table}

\subsubsection{Extension to complex null hypothesis}

\noindent In the previous sections, we considered an iterative SNP selection algorithm based on test statistics constructed from simple null hypotheses. In this section, we introduce a variant of the algorithm presented above, considering test statistics constructed from complex null hypotheses that involve several parameters at once. This complex-null extension is the joint Multiple-SNP construction underlying the NaM and ExM methods used throughout the paper, in which, from $k \ge 2$, the candidate SNP is tested \emph{jointly} with the already-selected set through an $F$-statistic rather than one coefficient at a time. The Multiple statistic is therefore a joint, prior-predictive evidence score for augmenting the current path with the candidate SNP rather than a classical partial test of the candidate coefficient alone that is the natural object for a stepwise scheme whose aim is to grow a well-supported set rather than to test one coefficient in isolation. The single-SNP Wald and Score variants of the previous sections and this joint variant are therefore two complementary configurations of the same algorithm. The two components that change from the algorithm presented above concern the definition and computation of the test statistic and of the posterior probabilities, detailed below.

\noindent For $k=1$, the new variant works exactly like the original version with the Wald statistic. For $k \ge 2$, one can rewrite model (\ref{modelk}) as 
\begin{equation}\label{model:multi}
Y = a_n \beta_{0jk} + \tilde X_{jk} \tilde \beta_{jk} + \varepsilon,
\end{equation}
where $\tilde X_{jk} = \begin{pmatrix}
X_{*j} & X^{(k)}
\end{pmatrix}$ is a $n \times k$ matrix, $\tilde\beta_{jk} = \begin{pmatrix}
\beta_{1jk} & \beta_{2jk}
\end{pmatrix}^{\top}$ is a vector of regression coefficients of dimension $k$ and $\varepsilon \sim \mathcal{N}(0, \sigma_{jk}^2I_n)$. Consider the null hypothesis $\tilde{\mathbb{H}}_0^{jk}: \tilde\beta_{jk} = \mathbf{0}_k$, where $\mathbf{0}_k$ is a null vector of $k$ dimension.
\noindent In Supplementary Appendix $C_1$, we show that the Wald statistic associated with the global null hypothesis $\tilde{\mathbb{H}}_0^{jk}: \tilde\beta_{jk} = \mathbf{0}_k$ is
$$\frac{Y^\top \tilde X_{jk}(\tilde X_{jk}^\top \tilde X_{jk})^{-1}\tilde X_{jk}^\top Y}{\sigma^2_{jk}}.$$

\noindent This statistic follows a non-central chi-squared distribution with $k$ degrees of freedom and a non-centrality parameter equal to $$\tilde\delta_{jk} = \tilde\Lambda_{jk}^\top \tilde  X_{jk}^\top \tilde X_{jk}\,\tilde\Lambda_{jk},$$
where $\tilde\Lambda_{jk}=\tilde\beta_{jk}/\sigma_{jk}=(\lambda_{jk},\lambda_{l_1},\lambda_{l_2},\cdots,\lambda_{l_{k-1}})^\top$ collects the standardized coefficients of model~(\ref{model:multi}).

\noindent When $\lambda_{jk}=0$, $\tilde\delta_{jk}$ reduces to
$$\delta_k = \Lambda_k^\top {X^{(k)}}^\top X^{(k)}\,\Lambda_k,$$
where $\Lambda_k = (\lambda_{l_1},\lambda_{l_2},\cdots,\lambda_{l_{k-1}})^\top$. We refer to the above approach as the naive multiple (NaM) method.

\noindent In practice, $\sigma^2_{jk}$ is unknown and is replaced by its REML estimate given by
$$\hat\sigma_{jk}^2 = \frac{1}{n-(k+1)} Y^\top \left\{I_n - \frac{1}{n} a_n a_n ^\top - \tilde X_{jk} (\tilde X_{jk}^\top \tilde X_{jk})^{-1} \tilde X_{jk}^\top\right\} Y.$$  

\noindent Thus, we obtain 

$$\tilde Z_{jk}^{(W)}=\frac{Y^\top \tilde X_{jk}(\tilde X_{jk}^\top \tilde X_{jk})^{-1}\tilde X_{jk}^\top Y/k}{\hat \sigma^2_{jk}}$$

\noindent and we show that $\tilde Z_{jk}^{(W)}$ follows a non-central Fisher distribution with $k$ and $n-(k+1)$ degrees of freedom and a non-centrality parameter equal to $\tilde\delta_{jk}$. Hereafter, we refer to this extension as the Exact Multiple (ExM) method. In practice, the NaM method uses the same plug-in estimate on the chi-squared scale, that is, it refers $k\,\tilde Z_{jk}^{(W)}$ to the non-central $\chi^2_k(\tilde\delta_{jk})$ law and ignores the variability of $\hat\sigma^2_{jk}$.

\noindent We define the stepwise posterior score $\tilde q_{jk}$ that SNP $j$ contributes to the phenotype as
$$
\tilde q_{jk} = P\!\left(\lambda_{jk} \neq 0 \,\Big|\, \tilde Z_{jk}^{(W)}=\tilde z_{jk}^{(W)}, \lambda_{l_1} \neq 0 , \lambda_{l_2} \neq 0, \cdots, \lambda_{l_{k-1}} \neq 0\right) = \frac{p_j\, \tilde A_{jk}}{p_j\, \tilde A_{jk} + (1-p_j)\, \tilde B_{jk}}.
$$
Here,
$$
\tilde A_{jk} = \int_{\delta=0}^\infty f_{k,n-(k+1)}\!\left(\tilde z_{jk}^{(W)};\delta\right) g_{jk}(\delta)\, d\delta
$$
is the conditional marginal density of $\tilde Z_{jk}^{(W)}$, $g_{jk}$ is the conditional prior density of $\tilde\delta_{jk}$ given $\{\lambda_{jk} \neq 0,\lambda_{l_1} \neq 0 , \cdots, \lambda_{l_{k-1}} \neq 0\}$, and $f_{k,n-(k+1)}(\cdot;\delta)$ is the density of the non-central Fisher distribution with $k$ and $n-(k+1)$ degrees of freedom and non-centrality parameter $\delta$.

\noindent Similarly,
$$\tilde B_{jk} = \int_{\delta=0}^\infty f_{k,n-(k+1)}\!\left(\tilde z_{jk}^{(W)};\delta\right) g_{k}(\delta)\, d\delta
$$
is the conditional marginal density of $\tilde Z_{jk}^{(W)}$ under the null on $\lambda_{jk}$, and $g_{k}$ is the conditional prior density of $\delta_{k}$ given $\{\lambda_{l_1} \neq 0 , \lambda_{l_2} \neq 0, \cdots, \lambda_{l_{k-1}} \neq 0\}$. \\

\noindent Because $\tilde\delta_{jk}$ and $\delta_k$ are quadratic forms in the Gaussian vectors $\tilde\Lambda_{jk}$ and $\Lambda_k$, their prior densities $g_{jk}$ and $g_k$ have no closed-form expression. The integrals $\tilde A_{jk}$ and $\tilde B_{jk}$, which are all the score requires, can nevertheless be computed exactly. The key observation is that, once the effects are averaged over their Gaussian prior, the quadratic form $Y^\top \tilde X_{jk}(\tilde X_{jk}^\top \tilde X_{jk})^{-1}\tilde X_{jk}^\top Y/\sigma^2_{jk}$ in the numerator of $\tilde Z_{jk}^{(W)}$ is distributed as a weighted sum of independent $\chi^2_1$ variables, with weights $1+\tau^2\mu_1,\dots,1+\tau^2\mu_k$, where $\mu_1,\dots,\mu_k$ are the eigenvalues of $\tilde X_{jk}^\top\tilde X_{jk}$. Therefore, its density is  a mixture of central Fisher densities with increasing numbers of degrees of freedom (Supplementary Appendix $C_2$):
$$
\tilde A_{jk} = \sum_{s\geq 0}\omega_s\,\frac{k}{k+2s}\,f_{k+2s,\,n-(k+1)}\!\left(\frac{k\,\tilde z_{jk}^{(W)}}{k+2s};\,0\right),
$$
where $f_{k+2s,\,n-(k+1)}(\cdot\,;0)$ is the density of the central Fisher distribution with $k+2s$ and $n-(k+1)$ degrees of freedom, and the mixing weights $\omega_s$ are the probabilities $P(N=s)$ of a sum $N$ of $k$ independent negative binomial variables with shape $1/2$ and probabilities $\theta_i = \tau^2\mu_i/(1+\tau^2\mu_i)$. The same expression gives $\tilde B_{jk}$ when the eigenvalues of $X^{(k)\top}X^{(k)}$ are used instead, and the NaM version is obtained by replacing the Fisher densities with the central chi-squared densities $f_{\chi^2_{k+2s}}(k\,\tilde z_{jk}^{(W)})$.

\noindent The weights $\omega_s$ follow from a short recursion whose terms are all positive, and the series is truncated with a guaranteed bound on the relative error. A rescaled form of the same series keeps the number of terms small for strong signals (Supplementary Appendix $C_2$). The stepwise posterior score $\tilde q_{jk}$ is thus computed exactly and deterministically, for all the candidates of a step at once, with no tuning parameter. Its cost is reported in Supplementary Table~\ref{tab:timing}. The joint scores use a common prior scale $\tau$ for the variants of the selected set.

\noindent A natural alternative would be to average $f_{k,n-(k+1)}(\tilde z_{jk}^{(W)};\tilde\delta_{jk})$ over draws of the effects from their prior. This Monte-Carlo average is unbiased, but it becomes unreliable when the prior spread of the non-centrality is large compared with the observed signal, as happens at large $n$, because only a few draws then fall where the density is non-negligible and these draws dominate the average. With $100$ draws per candidate, its relative error reached $40\%$ to $60\%$ at $n=1000$ in our simulations and significantly degraded the selection (Supplementary Tables~\ref{tab:exact_valid} and~\ref{tab:exact_select}). The exact computation removes this source of error, together with the need to choose a number of draws.

\subsubsection{Stopping criteria}

The iterative algorithm described above stops after $K$ steps. The number of steps can be predetermined by the user if the number of causal SNPs is assumed to be known. However, such an assumption is rarely defensible in practice when one analyzes a real dataset. In practice, the algorithm proceeds until a pre-specified stopping rule is met. In such a situation, $K$ becomes random. The stopping criteria considered in this work are {\it (i)} loss of adjusted-$R^2$, {\it (ii)} increase in BIC or in the extended BIC (eBIC) of \citet{chen2008} compared to the previous step, and {\it (iii)} maximum stepwise posterior score less than a threshold value specified by the user (we evaluate $0.5$, $0.95$ and $0.99$). The eBIC adds the penalty $2\gamma\log\binom{m}{k}$ to the standard BIC, which is particularly relevant in the high-dimensional setting $m \gg n$. In this work we set $\gamma=1$, the most conservative end of the usual range $\gamma \in [0,1]$. This is a parsimony setting suited to the present high-dimensional demonstration. A smaller $\gamma$ admits more variants and may be preferable for denser causal architectures, mirroring the architecture-dependence of the optimal stopping rule documented in Section~\ref{subsec:locus_sim}. These dynamic stopping rules terminate the variant selection and yield a set of SNPs that is parsimonious. These rules are practical implementation choices for terminating the greedy scan rather than part of the distributional theory.

\subsubsection{Use of the distributions beyond the prior-weighted score}\label{subsec:beyond}
\noindent Although we deploy the distributions of Table~\ref{tab:distributions} inside a prior-weighted stepwise score, they are not tied to that particular use. Because each distribution calibrates a candidate statistic \emph{conditional on the current selected set treated as fixed}, the same conditional Wald, Score and Multiple-SNP distributions can replace the nominal normal and chi-square references in a range of stepwise procedures. Examples include classical forward selection and conditional GWAS scans, SNP screening after conditioning on established lead variants, variable selection among correlated omics predictors, empirical-Bayes or Bayes factor ranking, and the calibration of otherwise ad hoc greedy algorithms. In each case the finite-sample correction sharpens the calibration of the per-step evidence, most visibly at small sample sizes. These distributions calibrate candidate statistics given a fixed conditioning set. 

\noindent Table~\ref{tab:methods_summary} summarizes our six methods and the external comparators used throughout the empirical sections.

\begin{table}[htbp] \centering \caption{Reading guide for the method variants. Proposed scores are stepwise scores.} \label{tab:methods_summary} \small \renewcommand{\arraystretch}{1.35} \setlength{\tabcolsep}{4pt} \begin{tabular}{ p{2.6cm} p{2.8cm} p{5.0cm} p{3.8cm} } \hline \textbf{Variant} & \textbf{Methods} & \textbf{Statistic/distribution} & \textbf{Use} \\ \hline Single-SNP normal & NaW, AsW, AsS & Wald/Score with naive or first-order normal & Fast screening, robust fallback. \\[2mm] Single-SNP exact & ExW & Wald with exact noncentral $t$ & Finite-sample single-SNP default. \\[2mm] Joint multiple & NaM, ExM & Candidate plus selected set through $F$-type score & Prior amplification under reliable annotation. \\[2mm] External & SuSiE, varbvs, LASSO, BayesC, CARMA, FINEMAP & Native scores, PIPs or selected sets & External assessments and locus-level checks. \\ \hline \end{tabular} \end{table}

\subsection{Comparative metrics of selection methods}

In this section, we define several metrics to assess the performance of the proposed iterative algorithm by simulations. Let $M \in \{0,1,\cdots,K\}$ be the number of true causal SNPs among the $K$ SNPs selected by the algorithm. The empirical estimates $E[M]$ and $E[K]$ are given by
$$\hat M = \frac{1}{B} \sum_{b=1}^B M_b \, \, \, \mbox{ and } \, \, \, \hat K = \frac{1}{B} \sum_{b=1}^B K_b,$$
where $B$ is the number of replications of the simulation study. Here, $M_b$ and $K_b$ are the values of $M$ and $K$ obtained from the $b^{\mbox{th}}$ generated dataset of the simulation study. The ratio $M/K$ represents the proportion of true positives. The empirical estimate of $\mu = E[M/K]$ is 
$$\hat\mu = \frac{1}{B} \sum_{b=1}^B \frac{M_b}{K_b}.$$

\noindent The tables also report the F1 score, the harmonic mean of the precision $M/K$ and the recall $M/q$, where $q$ is the number of causal SNPs. It equals $2M/(K+q)$, which is $0$ when no causal SNP is selected, including when the selection is empty, so that such a replication counts as a failure in the average over replications.

\noindent Consider now two competing selection methods and denote by $M^{(1)}$ and $M^{(2)}$ the numbers of true causal SNPs selected by these methods, respectively. Similarly, let $K^{(1)}$ and $K^{(2)}$ denote the numbers of steps run by the two algorithms. We compare their performances using the pairwise probability of superiority \citep{Rahardja2010}, defined by
$$\pi = \mathbb{P}\left(\frac{M^{(1)}}{K^{(1)}}>\frac{M^{(2)}}{K^{(2)}}\right) + \frac{1}{2} \times \mathbb{P}\left(\frac{M^{(1)}}{K^{(1)}}=\frac{M^{(2)}}{K^{(2)}}\right)$$

\noindent Through a simulation study with $B$ replicates, one may estimate $\pi$ by 

$$\hat\pi = \frac{1}{B} \sum_{b=1}^B 1_{\left\{\frac{M_b^{(1)}}{K_b^{(1)}}>\frac{M_b^{(2)}}{K_b^{(2)}}\right\}} + \frac{1}{2} \times \frac{1}{B} \sum_{b=1}^B 1_{\left\{\frac{M_b^{(1)}}{K_b^{(1)}}=\frac{M_b^{(2)}}{K_b^{(2)}}\right\}};$$
where $M^{(1)}_b/K_b^{(1)}$ and $M^{(2)}_b/K_b^{(2)}$ are the observed proportions of true causal SNPs (true positives) identified by the two competing methods from the same $b^{\mbox{th}}$ generated dataset ($b=1,\cdots,B$). Because both methods are applied to the same datasets, $\pi$ is a paired quantity. It is the probability that method~1 attains a higher proportion of true positives than method~2 on the same dataset, with ties counted as one half. A value $\pi > 0.5$ means that method~1 attains the higher proportion more often than method~2. For example, $\pi = 0.8$ means that method~1 does better in $80\%$ of the datasets when ties are counted as one half. A value $\pi = 0.5$ means that neither method tends to do better on the same dataset, and it does not imply that the two distributions of $M/K$ are equal. The estimator $\hat\pi$ is a sign-test-type statistic on the paired differences $M_b^{(1)}/K_b^{(1)}-M_b^{(2)}/K_b^{(2)}$. It differs from the unpaired probability of superiority, which compares every replicate of one method with every replicate of the other and equals the normalized Mann-Whitney $U$ statistic, and hence the area under the ROC curve \citep{Grissom1994,Ruscio2012}. A replication with an empty selection ($K=0$) contributes the proportion $0$, consistent with the F1 convention above. \\

\section{Simulations}

\subsection{Designs}\label{subsec:sim_designs}

Three complementary designs address distinct questions. The controlled AR(1) design isolates the effects of finite-sample calibration and prior quality. The single-locus design provides a directly comparable setting for locus-level fine-mapping methods. The semi-synthetic GEUVADIS design tests the same mechanisms under real human LD.

\paragraph{Controlled AR(1) design.}\label{subsec:locus_sim}
The controlled study uses $m=5{,}000$ SNPs in $100$ independent blocks of $50$, with within-block correlation $R_{ab}=\rho^{|a-b|}$ for $\rho\in\{0.8,0.95,0.98\}$ and $\mathrm{MAF}_j\sim\mathcal{U}(0.05,0.50)$. For each replicate,
$$Y_i=\sum_{j\in\mathcal C}\lambda X_{ij}+\epsilon_i, \qquad \epsilon_i\sim\mathcal N(0,1),$$
where $\mathcal C$ contains $15$ causal SNPs placed in distinct blocks. The common effect $\lambda$ is calibrated to marginal single-SNP power $\gamma\in\{0.2,0.6,0.9\}$ at $\alpha=0.05$. The main grid uses $n=1000$ with $B=50$ replications and the small-sample grid uses $n=80$ with the same number of replications. We consider $\tau_j\in\{0.2,0.5,1\}$ and four prior scenarios ranging from informative to misspecified, which are defined in Table~\ref{tab:parameters}. Because causal variants are known in simulation, the strongest informative prior is treated as an oracle reference rather than as a realistic annotation model. A separate graded-accuracy experiment then replaces this oracle structure by noisy annotations of known AUC.

\begin{table}[htbp]
\caption{\textbf{Prior scenarios in the controlled simulations.} The causal set is denoted by $\mathcal C$.}
\label{tab:parameters}
\centering\small
\begin{tabular}{ll}
\toprule
   & \textbf{Scenarios ($p_{j}$)}  \\
\midrule
\quad I. Informative   & $p_j = 0.8$ for $j \in \mathcal{C}$,\, $p_j = 0.2$ elsewhere. \\
\quad II. Uniform      & $p_j = 0.5$ for all $j$. \\
\quad III. Misspecified & $p_j = 0.2$ for $j \in \mathcal{C}$,\, $p_j = 0.5$ elsewhere. \\
\quad IV. Mixed        & $p_j \in \{0.2, 0.5, 0.8\}$ on $\mathcal{C}$ (five causals each),\, $p_j = 0.5$ elsewhere. \\
\bottomrule
\end{tabular}
\end{table}

\paragraph{Single-locus design for fine-mapping comparators.}
A single AR(1) locus with $m_{\mathrm{locus}}=200$ SNPs, $\rho=0.9$, three causal variants, $n=1000$ and $B=200$ replications is used for CARMA and FINEMAP. The remaining settings follow the controlled design.

\paragraph{Semi-synthetic GEUVADIS design.}
Thirty GEUVADIS cis windows provide real human genotype correlation structures. Five causal variants per window are sampled with probabilities proportional to PolyFun weights, which creates a truth-aligned but imperfect functional prior. Phenotypes are simulated at $\gamma=0.6$ and $\tau_j=0.2$ for $n\in\{80,373\}$ with $B=50$ replications per window and sample size.

\subsection{Results}\label{subsec:sim_results}

\paragraph{Prior quality and the relative behavior of single-SNP and joint scores.}
We first compare the proposed method families before turning to external competitors. At $n=1000$ and fixed $K=15$, the single-SNP methods are relatively robust to the prior, with recovery degrading gradually from the informative to the misspecified scenario and ExW and AsS leading NaW under the informative prior. The joint scores are much more sensitive to the annotation. NaM and ExM are nearly identical, approach the prior-only reference under the informative prior, and fall below the single-SNP scores once the prior is uniform or misspecified (Supplementary Table~\ref{tab:synth_scenarios}). These results show that the joint scores amplify prior information in both directions, while the single-SNP scores place more weight on the association evidence.

The graded-accuracy experiment makes this distinction explicit. Table~\ref{tab:noisy_prior} replaces the oracle prior by a noisy annotation whose AUC ranges from $0.60$ to $1.00$. When the annotation is imperfect, every proposed score improves mean recovery over the prior-only reference, which confirms that the data contribute to the ranking. The single-SNP scores remain ahead of the joint scores at every realistic accuracy level, and the large joint advantage appears only in the oracle limit. Joint scores are therefore best viewed as aggressive prior amplifiers, whereas ExW and AsS provide the more robust choice when annotation accuracy is uncertain.

\begin{table}[htbp]
\caption{\textbf{Graded prior accuracy.} Mean causal variants recovered out of $15$ with F1 in parentheses across annotation AUC values in the controlled AR(1) design. The first row gives the number of causal variants assigned $p_j=0.8$, and the prior-only row ranks by $p_j$ with marginal $|r|$ used to break ties.}
\label{tab:noisy_prior}
\centering\small
\begin{tabular}{p{3.5cm}cccc}
\toprule
Method & AUC $0.60$ & AUC $0.75$ & AUC $0.90$ & AUC $1.00$ \\
\midrule
\emph{Causal variants in the $p_j=0.8$ stratum} & 0.12 & 0.44 & 2.14 & 15.00 \\
\midrule
Prior-only ranking & 0.12 (0.008) & 0.44 (0.029) & 2.14 (0.143) & 15.00 (1.000) \\
AsS & 2.02 (0.135) & 2.20 (0.147) & 3.06 (0.204) & 9.12 (0.608) \\
ExW & 1.94 (0.129) & 2.30 (0.153) & 3.08 (0.205) & 9.20 (0.613) \\
NaM & 0.58 (0.039) & 0.88 (0.059) & 2.46 (0.164) & 14.82 (0.988) \\
ExM & 0.56 (0.037) & 0.88 (0.059) & 2.46 (0.164) & 14.82 (0.988) \\
\bottomrule
\end{tabular}
\end{table}

\paragraph{Finite-sample calibration.}
The small-sample experiment compares the distributional variants within each method family at $n=80$. ExW recovers substantially more causal SNPs than AsW at every power level in Table~\ref{tab:smalln}, and AsS closely matches ExW, which shows that estimating the variance under the null already removes most of the ranking loss of the asymptotic Wald construction. By contrast, ExM improves on NaM only marginally because all candidates at a given step share the same residual degrees of freedom and the Fisher correction changes their scores similarly.

\begin{table}[H]
\centering
\caption{\textbf{Small-sample fixed-$K$ results at $n=80$.} The obtained $\hat M$ (Mean causal variants recovered out of $15$) with F1 in parentheses under the informative prior.}
\label{tab:smalln}
\scriptsize
\setlength{\tabcolsep}{5pt}
\renewcommand{\arraystretch}{1.12}
\begin{tabular}{lccc}
\toprule
\textbf{Method} & $\gamma=0.2$ & $\gamma=0.6$ & $\gamma=0.9$ \\
\midrule
AsW & 2.61 (0.174) & 4.31 (0.287) & 5.83 (0.389) \\
ExW & 4.54 (0.303) & 6.94 (0.463) & 8.77 (0.584) \\
$\Delta$(ExW-AsW) & +1.93 (+0.129) & +2.63 (+0.176) & +2.93 (+0.196) \\
\addlinespace[2pt]
AsS & 4.59 (0.306) & 6.93 (0.462) & 9.22 (0.615) \\
\addlinespace[2pt]
NaM & 14.43 (0.962) & 14.66 (0.978) & 14.78 (0.986) \\
ExM & 14.56 (0.971) & 14.81 (0.987) & 14.91 (0.994) \\
$\Delta$(ExM-NaM) & +0.14 (+0.009) & +0.15 (+0.010) & +0.13 (+0.009) \\
\addlinespace[2pt]
SuSiE\textsubscript{prior} & 7.53 (0.502) & 8.35 (0.557) & 8.39 (0.559) \\
SuSiE\textsubscript{unif} & 0.27 (0.018) & 0.93 (0.062) & 1.45 (0.096) \\
varbvs & 0.27 (0.018) & 0.91 (0.061) & 1.63 (0.109) \\
\bottomrule
\end{tabular}
\end{table}

\begin{figure}[htbp]
\centering
\includegraphics[width=0.6\textwidth]{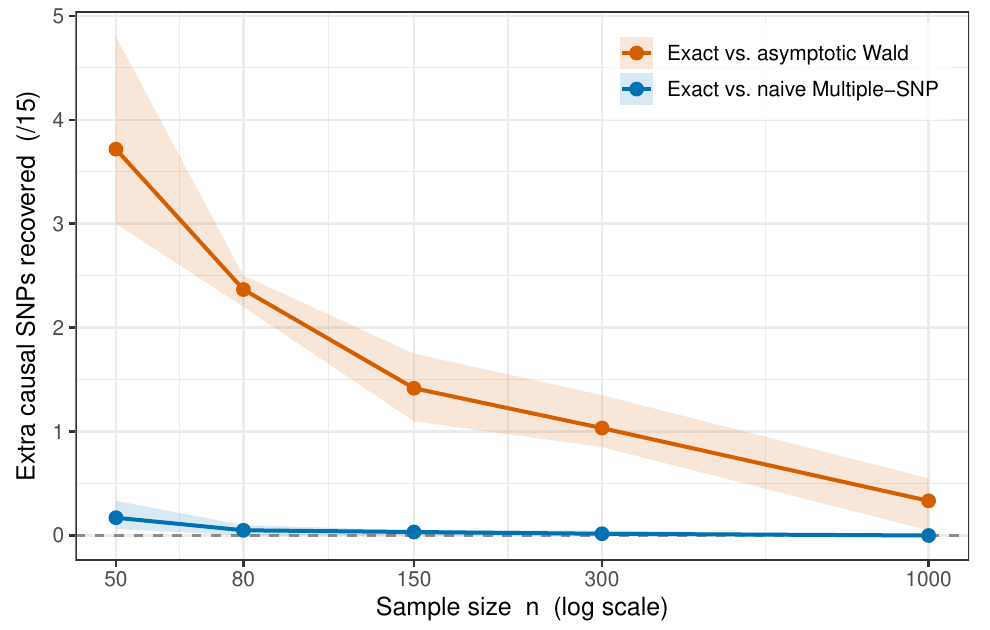}
\caption{\textbf{Finite-sample gain across sample size.} Extra causal variants recovered by ExW relative to AsW and by ExM relative to NaM in the controlled AR(1) design under the informative prior.}
\label{fig:premium_vs_n}
\end{figure}

Figure~\ref{fig:premium_vs_n} extends this comparison over sample size. Under the informative prior, the ExW advantage over AsW is largest at the smallest samples and nearly vanishes at $n=1000$, while the ExM advantage over NaM remains negligible throughout. The corresponding sweep under uniform and misspecified priors is essentially flat around zero in Supplementary Figure~\ref{fig:premium_by_prior}. Supplementary Table~\ref{tab:null_calibration} separates calibration from ranking by showing that the exact conditional law remains nominal for a fixed reference candidate under the global null. The finite-sample correction therefore improves the sampling calibration directly, while its effect on ranking is most visible when it changes the balance between association evidence and an informative prior.

\paragraph{Stopping rules.}
Stopping affects the method families differently. Under the prespecified BIC rule at $n=1000$, the joint scans select much shorter lists than the single-SNP scans and reach higher F1 values. BIC is retained as the default because it does not depend on the prior and gives an intermediate level of parsimony. Adjusted $R^2$ is more liberal, while eBIC is more conservative. Posterior-score stopping can recover the true model size under the oracle prior, but its behavior deteriorates sharply when the prior is uniform or misspecified. The complete comparison is reported in Supplementary Tables~\ref{tab:stop_joint}, \ref{tab:synth_informative_adaptive} and \ref{tab:synth_adaptive_best}.

\paragraph{External comparators in the single-locus design.}
The single-locus experiment provides the most direct comparison with fine-mapping methods. Under the informative prior and fixed $K=3$, NaM and ExM rank first, ExW follows, SuSiE with the matched prior comes next, and the prior-free CARMA and FINEMAP trail by a wide margin. Supplying CARMA with the annotation removes much of this separation, which confirms that the apparent gap is mainly attributable to prior information and not to the search algorithm alone. The four-scenario analysis further shows a robustness difference. Under the reversed annotation, CARMA$_p$ retains most of its recovery because it estimates the annotation coefficient, whereas NaM and ExM collapse because they take the supplied weights as fixed. Full fixed and adaptive results are reported in Supplementary Tables~\ref{tab:locus_sim_informative_fixed} to \ref{tab:locus_scenarios}.

The larger $m=5{,}000$ design is retained as a dimension stress test rather than as the primary fine-mapping comparison. SuSiE is run with $L=20$ effects, so its capacity exceeds the $15$ simulated causal variants. LASSO has near-zero recovery in this weak and distributed signal regime, whereas it performs strongly on the concentrated GEUVADIS eQTL signals of Section~\ref{sec:realdata}. This contrast is treated as a difference in operating regime rather than as evidence of general method superiority.

\paragraph{Real human LD.}\label{subsec:semisynth_main}
The semi-synthetic GEUVADIS results are reported at the LD-cluster level in Table~\ref{tab:geuvadis_semisynth}. Within the single-SNP family, ExW and AsS are essentially indistinguishable from NaW and AsW under this dense human LD, and within the joint family NaM and ExM are also nearly identical. With the truth-aligned PolyFun prior, the proposed methods tag more causal clusters than SuSiE at both sample sizes, and under the uniform prior they converge to a common intermediate level that still exceeds SuSiE. These results confirm that prior relevance transfers to real LD, while the practical difference between exact and approximate calibration is largely masked by the resolution limit imposed by dense correlation.

\begin{table}[htbp]
\caption{\textbf{Semi-synthetic GEUVADIS evaluation.} Mean causal LD clusters tagged out of $5$ across thirty cis windows under PolyFun and uniform priors.}
\label{tab:geuvadis_semisynth}
\centering\small
\begin{tabular}{lcccc}
\toprule
 & \multicolumn{2}{c}{$n=80$} & \multicolumn{2}{c}{$n=373$} \\
\cmidrule(lr){2-3}\cmidrule(lr){4-5}
Method & PolyFun & Uniform & PolyFun & Uniform \\
\midrule
NaW & 1.15 & 1.02 & 1.43 & 1.28 \\
AsW & 1.15 & 1.02 & 1.43 & 1.28 \\
ExW & 1.17 & 1.01 & 1.44 & 1.28 \\
AsS & 1.17 & 1.01 & 1.43 & 1.28 \\
NaM & 1.12 & 0.90 & 1.51 & 1.27 \\
ExM & 1.14 & 0.90 & 1.51 & 1.27 \\
SuSiE & 0.84 & 0.74 & 0.98 & 0.89 \\
\bottomrule
\end{tabular}
\end{table}

\paragraph{Computational cost.}
Runtime is reported in Supplementary Table~\ref{tab:timing}. The naive and asymptotic single-SNP scans require well under a second per GEUVADIS window at fixed $K=5$, whereas the exact variants pay for their prior integrals, with ExW the most expensive because it repeatedly evaluates non-central $t$ densities. Runtime scales approximately linearly with the number of candidates across the observed window sizes in Supplementary Figure~\ref{fig:scaling_m}. Thus the exact variants incur a measurable computational cost, but all reported GEUVADIS analyses remain below one minute per window on the reference workstation.

\section{Data analysis}\label{sec:realdata}

We illustrate the framework with a cis-eQTL prioritization analysis of human gene expression from the GEUVADIS study \citep{lappalainen2013}. GEUVADIS provides RNA-sequencing of lymphoblastoid cell lines (LCLs) and individual-level genotypes from the 1000~Genomes Project \citep{the1000genomes2015}. Each gene therefore defines a natural locus-level candidate set, and the analysis asks which variants within the cis window should be prioritized for follow-up.

\subsection{Data and design}\label{subsec:geuvadis_design}

We analyze the $462$ GEUVADIS individuals with both genotype and expression data. The $373$ European-ancestry samples (CEU, FIN, GBR and TSI) form the discovery set and the $89$ Yoruba (YRI) samples form a held-out replication panel at the association-analysis stage. We selected the $100$ genes with the strongest reported cis-association among the FDR-$5\%$ eGenes of the official European-only analysis (EUR373), restricted to chromosomes~1 to 7 because these chromosomes are covered by the functional-prior resource used below. Locus selection uses no YRI association information. Each candidate window spans $\pm1$~Mb around the transcription start site (TSS) and retains biallelic SNPs with discovery-set minor allele frequency $\geq0.05$, yielding $2{,}233$ to $20{,}097$ variants per window (median $4{,}424$).

\noindent Expression phenotypes are processed in three stages. We start from the official GEUVADIS gene-level RPKM quantifications, which were filtered, library-depth normalized and residualized on $10$ PEER factors by the consortium \citep{lappalainen2013}. Because this preprocessing predates our EUR/YRI split, the replication panel is held out at the association-analysis level rather than at the original expression-preprocessing stage. Each gene is then rank-inverse-normal transformed separately within EUR and YRI. Finally, phenotype and genotype columns are residualized on the fixed covariates used in each panel and include the intercept, sex and population indicators (FIN, GBR and TSI relative to CEU) in EUR, together with the intercept and sex in YRI. The residualized data are mapped to an orthonormal basis of the corresponding residual space, so the degrees of freedom used by the $t$ and $F$ laws are reduced by the exact rank of the projected-out covariate block. On these observed transformed phenotypes, the distributions of Section~2 are therefore interpreted as working-model exact laws, while formal exactness is reserved for the Gaussian simulations. Selection and replication statistics are computed from the residualized standardized data, whereas LD is computed from raw allele dosages within the corresponding population.

\subsection{Variant-level prior specifications}\label{subsec:geuvadis_priors}

Because the causal variants are unknown in observed data, the application varies the biological relevance rather than the unobservable fidelity of the prior. Four specifications are considered. The uniform prior is the neutral reference. The PolyFun prior is constructed from precomputed per-SNP heritability estimates from a UK Biobank multi-trait analysis \citep{weissbrod2020}, and it captures generic functional enrichment but is not specific to LCL cis-regulation. The TSS-proximity prior uses a well-established cis-eQTL enrichment feature, assigning greater weight to variants closer to the TSS. The discordant prior reverses the TSS weights and provides a deliberately adverse specification. All non-uniform priors use the three-level scale $p_j\in\{0.3,0.5,0.7\}$. The TSS prior assigns $0.7$ within $10$~kb, $0.5$ within $100$~kb and $0.3$ beyond, while the PolyFun prior assigns $0.3$ to zero-weight variants and splits positive weights at their within-window median, and the discordant prior is $1-p_j^{\mathrm{TSS}}$. The in-house methods and SuSiE receive the same $p_j$ within each arm. These four arms test whether the rankings benefit from a biologically relevant prior and how strongly they deteriorate when the prior is misleading.

\subsection{Cross-population evaluation}\label{subsec:geuvadis_replication}

True causal variants are not observed, so the primary evaluation uses cross-population support in YRI. Let $r_j^E$ and $r_j^Y$ be the marginal correlations between the residualized phenotype and variant $j$ in EUR and YRI. A YRI variant is considered associated when its effect is sign-consistent with EUR and its two-sided correlation $t$ statistic is significant. We report two replication thresholds, namely nominal $\alpha=0.05$ and a per-window permutation threshold given by the $95$th percentile of the maximal $|t|$ statistic over $500$ permutations of the residualized YRI phenotype. Because association data cannot distinguish near-equivalent proxies, the primary support metric is LD-aware, so an EUR-selected variant receives credit when it is in YRI LD ($|r|\geq0.7$ on raw dosages) with at least one YRI-associated variant, including itself. Thus, this metric measures cross-population support for the tagged signal rather than proof that the selected SNP itself is causal. Complementary analyses report exact-variant support, a Bonferroni threshold, conditional replication, out-of-sample adjusted $R^2$, and the number of distinct LD clusters represented by each selected set.

\subsection{Results}\label{subsec:geuvadis_results}

\paragraph{Single-SNP methods.}
At the common fixed budget $K=5$, NaW and AsW yield the same LD-aware support rates in every prior arm, while AsS is equal or slightly higher and ExW is lower on these data (Table~\ref{tab:geuvadis_fixedK}). The clearest prior effect within this family is the TSS-proximity gain, which raises the nominal support rate of every single-SNP score relative to the uniform prior, and the block-bootstrap intervals for the TSS-minus-uniform difference exclude zero for all four scores (Supplementary Table~\ref{tab:geuvadis_ci}). PolyFun changes the single-SNP rates only marginally. The permutation-controlled analysis preserves the same broad conclusion, namely a TSS benefit for the single-SNP scores, without requiring the exact ordering of all four priors to be identical. In this $n=373$ application, ExW therefore provides no empirical ranking advantage over AsS, consistent with the simulation result that the finite-sample Wald gain is concentrated at substantially smaller sample sizes.

\paragraph{Joint methods.}
NaM and ExM are practically indistinguishable in the real-data rankings, with nominal and permutation-controlled fixed-$K$ rates differing by at most $0.006$ across prior arms. In contrast to the single-SNP family, neither PolyFun nor the TSS prior improves the joint scores over the uniform reference, and the TSS-minus-uniform block-bootstrap contrasts are slightly negative (Supplementary Table~\ref{tab:geuvadis_ci}). The dominant prior effect is instead the penalty from the discordant arm, which markedly lowers both the nominal and the permutation-controlled rates. Thus, on these strong cis-eQTLs, the joint scores mainly reveal their prior-amplifying behavior through their sensitivity to a misleading annotation. The prior-amplitude analysis leads to the same conclusion (Supplementary Table~\ref{tab:geuvadis_prior_scale}).

\paragraph{Single-SNP versus joint selection.}
The two families optimize different aspects of prioritization. At fixed $K=5$, the joint scores have much higher per-variant LD-aware support than the single-SNP scores in the non-discordant arms (Table~\ref{tab:geuvadis_fixedK}). This advantage reflects a markedly more concentrated selection structure, since the five NaM/ExM variants cover about two distinct EUR LD clusters on average whereas the single-SNP scans spread over nearly five (Supplementary Table~\ref{tab:geuvadis_clusters}). Once each LD cluster is counted only once, the single-SNP scans recover approximately $1.5$ to $1.7$ distinct supported signals per gene, compared with about $1.2$ to $1.4$ for the joint scores. The per-variant support rate and the number of distinct signals should therefore be interpreted jointly because the joint scores favor compact concentration on a small number of strongly supported signals, whereas the conditional single-SNP scores provide broader signal coverage.

\begin{table}[htbp]
\caption{\textbf{Fixed-budget cross-population support in GEUVADIS.} All methods in this table select $K=5$ variants per gene. Each cell gives the mean LD-aware support rate over the $100$ genes under the nominal YRI test, followed by the rate under the per-window permutation threshold.}
\label{tab:geuvadis_fixedK}
\centering\small
\setlength{\tabcolsep}{4pt}
\begin{tabular}{lcccc}
\toprule
Method & Uniform & PolyFun & TSS & Discordant \\
\midrule
\multicolumn{5}{l}{\emph{Single-SNP stepwise scores}} \\
NaW & 0.332 / 0.174 & 0.334 / 0.178 & 0.396 / 0.210 & 0.320 / 0.172 \\
AsW & 0.332 / 0.174 & 0.334 / 0.178 & 0.396 / 0.210 & 0.320 / 0.172 \\
ExW & 0.308 / 0.158 & 0.312 / 0.150 & 0.354 / 0.174 & 0.298 / 0.150 \\
AsS & 0.332 / 0.172 & 0.338 / 0.182 & 0.408 / 0.216 & 0.324 / 0.174 \\
\addlinespace
\multicolumn{5}{l}{\emph{Joint stepwise scores}} \\
NaM & 0.780 / 0.536 & 0.720 / 0.482 & 0.740 / 0.496 & 0.562 / 0.350 \\
ExM & 0.778 / 0.536 & 0.718 / 0.482 & 0.734 / 0.496 & 0.566 / 0.344 \\
\addlinespace
\multicolumn{5}{l}{\emph{Fine-mapping comparator at the same budget}} \\
SuSiE & 0.706 / 0.470 & 0.702 / 0.452 & 0.714 / 0.470 & 0.708 / 0.458 \\
\bottomrule
\end{tabular}
\end{table}

\paragraph{Comparison with external methods.}
At the same fixed budget and with the same prior weights, the joint scores are modestly above SuSiE in the three non-adversarial arms but are substantially worse under the discordant prior, where SuSiE is nearly unaffected (Table~\ref{tab:geuvadis_fixedK}). The matched-budget sensitivity likewise places NaM/ExM above the top-PIP SuSiE ranking for $K'=1,\ldots,5$ under the TSS prior, while the prior-free LASSO remains the strongest per-variant evaluation over these budgets (Supplementary Figure~\ref{fig:budget_curve}). These results favor no universal winner because the proposed joint score gains when its external annotation is useful, but pays a larger price than SuSiE when that annotation is deliberately reversed. A paired per-gene probability-of-superiority analysis gives the same broad ordering, with row means of $0.74$ for LASSO, $0.68$ for NaM and SuSiE, $0.67$ for ExM and $0.25$ to $0.34$ for the single-SNP scores.

\paragraph{Adaptive and native operating points.}
The comparison is most natural when each procedure is allowed to return a sparse list. Under BIC stopping, NaM and ExM select fewer than two variants per gene and attain the highest nominal LD-aware support in every arm, with the discordant prior remaining the weakest (Table~\ref{tab:geuvadis_adaptive}). QTLtools, the field-standard conditional cis scan, selects a similarly short list on the same windows but reaches a lower support rate, so the BIC-stopped joint scores have higher mean per-variant support at a closely matched selection size. The permutation-controlled analysis leads to the same ordering, so the lead of the compact joint shortlists does not depend on the nominal replication threshold (Table~\ref{tab:geuvadis_perm_bic}). The one-standard-error LASSO selects a longer list with support close to QTLtools and has the largest out-of-sample adjusted $R^2$, whereas the BIC-stopped joint scores yield smaller $R^2$ values because they deliberately stop after a much shorter list (Supplementary Table~\ref{tab:geuvadis_r2}). By comparison, BIC-stopped single-SNP scans retain far longer lists, achieve lower per-variant support, and reach higher $R^2$ than the joint shortlists. Thus, the adaptive analysis reinforces the distinction between compact prioritization and broader predictive or multi-signal coverage.

The sensitivity analyses support this interpretation. Exact-variant and Bonferroni-controlled support preserve the principal method-family contrasts (Supplementary Table~\ref{tab:geuvadis_sensitivity}). Conditional replication is much harder in the smaller YRI panel and reduces all fixed-$K$ stepwise rates to a uniformly low level (Supplementary Table~\ref{tab:geuvadis_cond}), which confirms that the primary LD-aware metric should be interpreted as support for tagged signals. Finally, the semi-synthetic GEUVADIS evaluation of Section~\ref{subsec:semisynth_main}, where the causal truth is known, reaches the complementary conclusion that dense human LD limits exact-variant resolution and makes cluster-level recovery the more appropriate target.

\begin{table}[htbp]
\caption{\textbf{Adaptive and native sparse selections in GEUVADIS.} For the in-house methods, each cell reports the nominal LD-aware support rate, with the mean BIC-selected size $\hat K$ in parentheses.}
\label{tab:geuvadis_adaptive}
\centering\small
\begin{tabular}{lcccc}
\toprule
Method & Uniform & PolyFun & TSS & Discordant \\
\midrule
\multicolumn{5}{l}{\emph{BIC-stopped single-SNP scores}} \\
NaW & 0.225 (15.6) & 0.274 (10.7) & 0.331 (8.3) & 0.209 (14.6) \\
AsW & 0.224 (15.5) & 0.268 (10.6) & 0.334 (8.5) & 0.209 (14.4) \\
ExW & 0.207 (13.4) & 0.243 (9.3) & 0.282 (8.7) & 0.198 (13.2) \\
AsS & 0.224 (15.5) & 0.275 (10.5) & 0.344 (8.2) & 0.209 (14.3) \\
\addlinespace
\multicolumn{5}{l}{\emph{BIC-stopped joint scores}} \\
NaM & 0.852 (1.8) & 0.840 (1.5) & 0.830 (1.4) & 0.777 (1.5) \\
ExM & 0.852 (1.8) & 0.845 (1.5) & 0.830 (1.4) & 0.783 (1.5) \\
\midrule
LASSO (1-SE, prior-free) & \multicolumn{4}{c}{0.792 \quad ($\hat K=3.8$)} \\
QTLtools (prior-free) & \multicolumn{4}{c}{0.792 \quad ($\hat K=1.5$)} \\
\bottomrule
\end{tabular}
\end{table}

\begin{table}[htbp]
\caption{\textbf{Permutation-controlled replication of the BIC-stopped selections.} The compact joint shortlists keep their lead, so their replication performance does not depend on the nominal $\alpha=0.05$ threshold.}
\label{tab:geuvadis_perm_bic}
\centering\small
\begin{tabular}{lcccc}
\toprule
Method & Uniform & PolyFun & TSS & Discordant \\
\midrule
NaW & 0.088 (15.6) & 0.130 (10.7) & 0.163 (8.3) & 0.090 (14.6) \\
AsW & 0.089 (15.5) & 0.130 (10.6) & 0.165 (8.5) & 0.090 (14.4) \\
ExW & 0.085 (13.4) & 0.115 (9.3) & 0.127 (8.7) & 0.080 (13.2) \\
AsS & 0.088 (15.5) & 0.132 (10.5) & 0.175 (8.2) & 0.092 (14.3) \\
NaM & 0.586 (1.8) & 0.563 (1.5) & 0.591 (1.4) & 0.547 (1.5) \\
ExM & 0.588 (1.8) & 0.567 (1.5) & 0.591 (1.4) & 0.552 (1.5) \\
\midrule
LASSO (1-SE, prior-free) & \multicolumn{4}{c}{0.513 \quad ($\hat K=3.8$)} \\
QTLtools (prior-free) & \multicolumn{4}{c}{0.518 \quad ($\hat K=1.5$)} \\
\bottomrule
\end{tabular}
\end{table}

\clearpage

\section{Discussion}\label{sec:discussion}

We proposed a prior-informed stepwise framework for SNP prioritization based on finite-sample and approximate association-test distributions. The exact variants use the non-central $t$ and $F$ laws induced by the Gaussian linear working model, while the asymptotic variants replace the estimated residual variance by first-order approximations. The simulation controls clarify what this correction changes. For a fixed candidate under the global null, the exact law is correctly calibrated, while the normal reference shows mild small-sample distortion. At the ranking level, however, the practical ExW advantage over AsW is concentrated in the informative-prior setting because the exact law changes how strong observed signals are balanced against prior weights. AsS reaches nearly the same small-sample recovery as ExW, and ExM and NaM select almost the same variants. The exact distributional correction is therefore most consequential for Wald-based single-SNP scoring at small or moderate sample size, while the joint exact law mainly provides principled calibration rather than a large ranking advantage.

\begin{table}[htbp] \centering \caption{Practical recommendation.} \label{tab:recommendations} \small \renewcommand{\arraystretch}{1.35} \setlength{\tabcolsep}{4pt} \begin{tabular}{ >{\raggedright\arraybackslash}p{2.9cm} >{\raggedright\arraybackslash}p{3.9cm} >{\centering\arraybackslash}p{1.6cm} >{\raggedright\arraybackslash}p{5.6cm} } \hline \textbf{Annotation setting} & \textbf{Preferred method} & \textbf{Stopping} & \textbf{Rationale} \\ \hline Externally validated or very high-fidelity prior & NaM or ExM & BIC & Exploit reliable annotation. An AUC of 0.9 is not yet sufficient. \\[1.5mm] Weak/unknown prior & ExW or AsS & BIC/eBIC & Robust single-SNP discovery. \\[1.5mm] Possibly wrong prior & Avoid joint scans, use ExW/AsS & BIC/eBIC & Prevent harmful prior amplification. \\[1.5mm] Strong concentrated signals, no usable prior & LASSO or the native conditional scan & Native & Competitive baselines. The framework adds prior integration and calibrated conditional evidence. \\[1.5mm] Locus-level reporting & Proposed ranking + SuSiE/\allowbreak CARMA/\allowbreak FINEMAP & Native/\allowbreak local & Add credible-set-style context. \\[1.5mm] Small or moderate $n$ & ExW or AsS & BIC & The exact Wald law, or the Score statistic with null variance, corrects the small-sample miscalibration of the asymptotic Wald test. \\ \hline \end{tabular} \end{table}

\noindent The second simulation lesson concerns prior quality. The oracle prior is useful only as a reference because it identifies the causal set by construction. Once the annotation is degraded continuously, the method families separate more clearly. ExW and AsS remain relatively robust, while NaM and ExM behave as aggressive prior amplifiers. They perform exceptionally well only when the annotation is close to oracle quality and become inferior to the single-SNP scores once annotation accuracy falls. This behavior is also visible in the four-scenario experiment, where the joint scores track the informative and mixed priors closely but deteriorate sharply under the reversed annotation. The practical implication is that a non-uniform prior should not automatically trigger the joint mode. The joint score is appropriate when the annotation has been externally validated at high resolution, while single-SNP scoring is safer when prior quality is uncertain.

\noindent The stopping-rule analyses reinforce the same distinction. BIC provides a prior-independent intermediate level of parsimony and is therefore used as the default. Posterior-score thresholds can recover the true model size under an oracle prior, but their behavior depends strongly on the scale and direction of the annotation. Adjusted $R^2$ is more liberal and eBIC is more conservative. These rules are implementation choices for terminating the greedy path and do not provide formal false-discovery control.

\noindent The external comparisons are most informative in the single-locus design. ExM and NaM perform strongly under the informative prior, while prior-informed SuSiE and CARMA also improve when supplied with the same information. CARMA$_p$ is particularly informative because it estimates the annotation coefficient and can therefore recover from a reversed annotation, unlike the proposed scores, which take $p_j$ as fixed. This comparison shifts the emphasis away from universal method ranking and toward the interaction between prior quality and the way each method uses that prior. The larger $m=5{,}000$ design serves mainly as a dimensional stress test. The weak and distributed signal regime is unfavorable to the LASSO, whereas the strong and concentrated GEUVADIS eQTL signals are favorable to sparse penalized selection.

\noindent The semi-synthetic GEUVADIS experiment shows that these conclusions persist under real human LD. The truth-aligned PolyFun prior improves recovery at small sample size, while exact and approximate variants within each method family are nearly indistinguishable on the densely correlated windows. This suggests that, once LD imposes a strong resolution limit, improving the calibration of the candidate statistic cannot by itself recover information that is absent from the genotype correlation structure.

\noindent The GEUVADIS cis-eQTL application clarifies how these mechanisms translate to observed human data. Within the single-SNP family, TSS proximity produces a reproducible gain over the uniform prior, whereas the generic PolyFun prior changes the rankings little. The joint NaM and ExM scores are nearly indistinguishable and show a different pattern because their support is already high under the uniform prior, does not improve under TSS and falls markedly under the deliberately discordant prior. This is consistent with the simulation finding that the joint scores act as aggressive prior amplifiers rather than uniformly benefiting from any non-uniform annotation. The fixed-budget analysis also reveals a structural trade-off. Joint scores, SuSiE and the LASSO concentrate on a small number of highly correlated variants and obtain high per-variant cross-population support, whereas the conditional single-SNP scans cover substantially more LD clusters and slightly more distinct supported signals. Thus, per-variant support and signal coverage answer different prioritization questions and should not be conflated.

\noindent The external comparisons reinforce this interpretation rather than establishing a universal ranking of methods. At the same fixed budget, NaM and ExM are modestly above SuSiE under non-adversarial priors but fall well below it when the prior is reversed. At native sparse operating points, QTLtools selects about $1.5$ signals per gene and the LASSO about $3.8$ variants, and both achieve strong cross-population support. BIC-stopped NaM and ExM select about $1.4$ to $1.8$ variants and have numerically higher mean per-variant support than QTLtools at a similar selection size, while the LASSO attains the largest out-of-sample $R^2$. These differences reflect distinct objectives because the proposed joint scores use external annotation to construct a compact shortlist, QTLtools targets independent conditional cis signals and the LASSO optimizes a sparse penalized fit without incorporating variant-level prior information. No single procedure dominates simultaneously in per-variant support, distinct-signal coverage and out-of-sample prediction.

\noindent Two qualifications remain important. The application uses the hundred strongest EUR eGenes, for which leading signals are well resolved, and weaker loci may produce a different balance between primary and secondary-signal recovery. The YRI panel is also small with $n=89$. Conditional replication therefore has limited power beyond the leading signal and substantially attenuates differences among the stepwise methods. The primary LD-aware metric should consequently be interpreted as cross-population support for tagged signals rather than as independent confirmation of every conditional effect or of SNP-level causality.

\noindent Table~\ref{tab:recommendations} translates these findings into a practical workflow.

\noindent Several limitations remain. First, the present work treats $p_j$ and $\tau_j$ as fixed inputs. In practice, these weights require estimation, calibration or learning from external annotations, functional data or empirical Bayes procedures. Learning the weight of an annotation, as CARMA does, would also protect the joint scores against a reversed prior. Second, the algorithm is greedy and does not average over the full model space, so its stepwise posterior scores are local path-conditional quantities rather than global PIPs. Third, the exact test distributions are exact only under the stated Gaussian homoscedastic linear model. Relatedness, population structure and non-Gaussian phenotypes require covariates, residualization or mixed-model extensions. Fourth, in very high LD, no method can guarantee single-variant resolution from association data alone, so selected variants often act as representatives of LD clusters.

\noindent In summary, the method provides a computationally light bridge between marginal GWAS screening and full locus-level fine-mapping. It is strongest when used as a prior-sensitive prioritization tool. Single-SNP exact and score-based variants provide robust discovery, while joint variants provide concentrated prioritization when the biological prior is credible.

\section*{Data and code availability}

The GEUVADIS genotype and gene-expression data are publicly available from the EBI ArrayExpress/BioStudies repository (accession E-GEUV-1) \citep{lappalainen2013}, and the PolyFun per-SNP prior weights from the PolyFun repository \citep{weissbrod2020}. All numerical summaries of the main text and Supplement are reproducible from the public repository \href{https://github.com/Doulabe/exact-stepwise-finemap}{\texttt{github.com/\allowbreak Doulabe/\allowbreak exact-stepwise-finemap}}, which provides the simulation and real-data analysis code with fixed random seeds, the table and figure generation scripts, the preparation scripts assembling genotypes, expression phenotypes and prior weights, a README describing how to reproduce each main figure, and the computational environment (R and package versions).




\section*{Conflict of interest}

The authors declare no competing interests.



\bibliography{references}

\clearpage
\fancyhead[L]{Supplementary Material}
\setcounter{section}{0}\setcounter{subsection}{0}
\setcounter{table}{0}\setcounter{figure}{0}\setcounter{equation}{0}
\renewcommand{\thesection}{S\arabic{section}}
\renewcommand{\thetable}{S\arabic{table}}
\renewcommand{\thefigure}{S\arabic{figure}}
\renewcommand{\theequation}{S\arabic{equation}}
\makeatletter
\setlength{\@fptop}{0pt}
\setlength{\@fpsep}{12\p@ \@plus 0fil}
\setlength{\@fpbot}{0pt \@plus 1fil}
\makeatother
\renewcommand{\floatpagefraction}{0.85}
\renewcommand{\topfraction}{0.92}
\renewcommand{\bottomfraction}{0.9}
\renewcommand{\textfraction}{0.08}
\begin{center}
{\LARGE\bfseries Supplementary Material}\\[1ex]
{\large Exact association-test distributions for prior-informed stepwise SNP prioritization}
\end{center}
\noindent This supplementary material contains the mathematical appendix (derivations of the
test statistics and their distributions) and the supplementary tables and
figures referenced in the main text. Cross-references of the form ``Table~1'',
``Figure~2'' or ``Section~3'' refer to the main paper; supplementary items are
prefixed with ``S'' (Table~S1, Figure~S1, \ldots).

\section*{Appendix}
\subsection*{A1: Statistics $Z_{j1}^{(W)}$ and $Z_{j1}^{(S)}$}

Under the model $Y = a_n \beta_{0j1} + X_{*j}\beta_{1j1} + \varepsilon_{j1}$, with $a_n^\top a_n = X_{*j}^\top X_{*j} = n$ and $X_{*j}^\top a_n = 0$, the estimator $\hat{\beta}_{1j1} = X_{*j}^\top Y /n$ has $\widehat{Var}(\hat{\beta}_{1j1}) = n^{-1} \hat{\sigma}^2_{j1}$, so the Wald statistic is
$$Z_{j1}^{(W)} = \frac{\hat{\beta}_{1j1}}{\sqrt{\widehat{Var}(\hat{\beta}_{1j1})}} = \frac{X_{*j}^\top Y}{\sqrt{n\hat{\sigma}^2_{j1}}}.$$
\noindent Differentiating the Gaussian log-likelihood at $\beta_{1j1}=0$ gives the score $S = X_{*j}^\top (Y - a_n\beta_{0j1})/\sigma^2_{j1}$ and information $U = n/\sigma^2_{j1}$, so that, using $X_{*j}^\top a_n = 0$ and the null variance estimate,
\begin{equation*}
Z_{j1}^{(S)} = \left.\frac{S}{\sqrt{U}}\right|_{\beta_{1j1} = 0} = \frac{X_{*j}^\top Y}{\sqrt{n \tilde\sigma^2_{j1}}}.
\end{equation*}

\subsection*{A2: Distributions of $Z_{j1}^{(W)}$ and $Z_{j1}^{(S)}$}

\noindent\textbf{Asymptotic Wald law.} Write $Z_{j1}^{(W)} = g(\hat{\beta}_{j1}, \hat{\sigma}^2_{j1})$ with $g(a, b) = a/\sqrt{b/n}$. A first-order Taylor expansion of $g$ around $(\beta_{j1}, \sigma^2_{j1})$, with $\partial g/\partial a = \sqrt{n/\sigma^2_{j1}}$, $\partial g/\partial b = -\sqrt{n}\,\beta_{j1}/\{2 (\sigma^2_{j1})^{3/2}\}$ and $\text{Var}(\hat{\sigma}_{j1}^2) \approx 2 (\sigma^2_{j1})^2/n$, gives $\mathbb{E}(Z_{j1}^{(W)}) = \sqrt{n} \lambda_{j1}$ and $\mathbb{V}ar(Z_{j1}^{(W)}) = 1 + \lambda^2_{j1}/2$, where $\lambda_{j1} = \beta_{j1}/\sigma_{j1}$, so that
\begin{equation*}
    Z_{j1}^{(W)} \sim \mathcal{N}\left(\sqrt{n} \lambda_{j1}, \,  1 + \frac{1}{2} \lambda^2_{j1} \right).
\end{equation*}

\noindent\textbf{Asymptotic Score law.} With $P_{j1} = I_n - a_n a_n^\top/n$ (idempotent), $\tilde\sigma^2_{j1} = Y^\top P_{j1} Y/(n-1) \approx Y^\top P_{j1} Y/n$, so that $Z_{j1}^{(S)}/\sqrt{n} \approx g(\hat{a}, \hat{b})$ with $g(a,b) = a/\sqrt{b}$, $\hat{a} = X_{*j}^\top Y/(n\sigma_{j1})$ and $\hat{b} = Y^\top P_{j1}Y/(n\sigma^2_{j1})$. Direct computation gives $\hat{a} \sim \mathcal{N}(\lambda_{j1}, 1/n)$; by Corollary~2 of Theorem~5.5 in \citet{Chapter5Rencher}, $n\hat{b} \sim \chi^2(n-1, n\lambda_{j1}^2/2)$, so that $\mathbb{E}[\hat{b}]\approx 1 + \lambda_{j1}^2$ and $\mathbb{V}ar(\hat{b}) \approx 2(1+2\lambda_{j1}^2)/n$; and by Theorem~5.2.d of the same reference, $\text{Cov}(\hat{a}, \hat{b}) = 2\lambda_{j1}/n$. The first-order expansion of $g$ around $(a, b) = (\lambda_{j1}, 1+\lambda_{j1}^2)$ then yields $\mathbb{E}[Z_{j1}^{(S)}] = \sqrt{n}\lambda_{j1}/\sqrt{1 + \lambda_{j1}^2}$ and, collecting the three variance terms,
\begin{equation*}
\mathbb{V}ar(Z_{j1}^{(S)}) = \frac{1}{1 +\lambda_{j1}^2} + \frac{\lambda_{j1}^2(1+2\lambda_{j1}^2)}{2(1 +\lambda_{j1}^2)^{3}} - \frac{2\lambda_{j1}^2}{(1 +\lambda_{j1}^2)^{2}} = \frac{1 + \lambda_{j1}^2/2}{(1 +\lambda_{j1}^2)^3},
\end{equation*}
so that, asymptotically, $Z_{j1}^{(S)} \sim \mathcal{N} \left(\displaystyle \frac{\sqrt{n}\lambda_{j1}}{\sqrt{1 + \lambda_{j1}^2}}, \,  \frac{1}{(1 +\lambda_{j1}^2)^3}\left(1 + \frac{1}{2}\lambda_{j1}^2\right) \right)$.

\noindent These normal approximations for the Naive and Asymptotic Wald and Score statistics rest on first-order Taylor expansions, and are accurate when $n$ is large and the per-SNP signal $\lambda_{j1}$ is bounded away from the degenerate regime. The \emph{exact} Wald and Score variants do not rely on these expansions: conditionally on the standardized design, the non-central $t$ and $F$ laws derived above involve no approximation that depends on the allele frequency, and they are the recommended choice in the small-$n$ regime. At very low minor allele frequencies and small $n$, a genotype column takes few distinct values and its standardization becomes unstable, which affects all methods equally. Our analyses are restricted to common variants throughout (the synthetic design draws $\mathrm{MAF}_j \sim \mathcal{U}(0.05, 0.50)$ and the GEUVADIS analyses filter at $\mathrm{MAF} \geq 0.05$); this is a design restriction of the empirical studies, not a requirement of the theory.

\noindent In addition, we know that $\hat{\beta}_{1j1} \sim \mathcal{N}(\beta_{1j1}, \sigma^2_{j1}/n)$ and $\displaystyle\frac{(n-2) \hat{\sigma}^2_{j1}}{\sigma^2_{j1}} \sim \chi^2_{n-2}$. Similarly, one can see that $\sqrt{n}\hat{\beta}_{1j1} \sim \mathcal{N}(\sqrt{n} \beta_{1j1}, \sigma^2_{j1})$ and $\hat{\beta}_{1j1}$ and $\hat{\sigma}^2_{j1}$ are independent. As described in section 5.4.2 in \citet{Chapter5Rencher}, it is straightforward that the Wald statistic follows a non-central student distribution,
$$ Z_{j1}^{(W)} = \frac{\sqrt{n} \hat{\beta}_{1j1}/  \sigma_{j1}}{\sqrt{\frac{(n-2) \hat{\sigma}^2_{j1}}{\sigma^2_{j1}} / (n-2)}} \sim t(n-2, \, \sqrt{n}\beta_{1j1}/ \sigma_{j1}) = t(n-2, \, \lambda_{j1}\sqrt{n}).$$
This yields the one-to-one relationship mentioned in Section~\ref{sect2.2.2}.

\subsection*{B1 : Statistics $Z_{jk}^{(W)}$ and $Z_{jk}^{(S)}$}
Under the model $Y = a_n \beta_{0jk} + X_{*j}\beta_{1jk} + X^{(k)} \beta_{2jk} + \varepsilon_{jk}$,
 and $X^{(k)}$, $r_{jk}$, $W_k$ and $\rho_{jk}$ as already defined above in the text, if we define $\tilde{X} = \begin{pmatrix}
a_n & X^{(k)} & X_{*j}
\end{pmatrix}$, then $$\tilde{X} ^\top \tilde{X}  = n \begin{pmatrix}
1 & \mathbf{0} & \mathbf{0} \\
\mathbf{0} & W_k & r_{jk} \\
\mathbf{0} & r_{jk}^\top & 1
\end{pmatrix} \hspace{.2cm} \text{ and } \hspace{.2cm}
[\tilde{X} ^\top \tilde{X}]^{-1}  = \frac{1}{n}\begin{pmatrix}
1 & \mathbf{0} & \mathbf{0} \\
\mathbf{0} & \displaystyle W_k^{-1} + \frac{W_k^{-1} r_{jk} r_{jk}^\top W_k^{-1}}{1 - \rho_{jk}} & - \displaystyle \frac{W_k^{-1} r_{jk}}{1 - \rho_{jk}} \\
\mathbf{0} & - \displaystyle \frac{r_{jk}^\top W_k^{-1}}{1 - \rho_{jk}} & \displaystyle \frac{1}{1 - \rho_{jk}}
\end{pmatrix}.$$ 

\noindent The regression coefficients can now be estimated as
\begin{align*}
\hat{\beta} = \begin{pmatrix}
\hat{\beta}_{0jk} & \hat{\beta}_{2jk} & \hat{\beta}_{1jk}
\end{pmatrix}^\top & = [\tilde{X} ^\top \tilde{X}]^{-1} \tilde{X} ^\top Y \\
& = \frac{1}{n} \begin{pmatrix}
a_n^\top \\ \left(W_k^{-1} + \displaystyle\frac{W_k^{-1} r_{jk} r_{jk}^\top W_k^{-1}}{1 - \rho_{jk}}\right) X^{(k)\top} - \displaystyle \frac{W_k^{-1} r_{jk}}{1 - \rho_{jk}} X_{*j}^\top \\  \displaystyle \frac{1}{1 - \rho_{jk}} \left(X_{*j}^\top - r_{jk}^\top W_k^{-1} X^{(k)\top}\right)
\end{pmatrix} Y.
\end{align*}

\noindent From the above equation, $\hat{\beta}_{1jk} = \displaystyle \frac{\left(X_{*j}^\top - r_{jk}^\top W_k^{-1} X^{(k)\top}\right)Y}{n(1 - \rho_{jk})} $. Since $\mathbb{E}[\hat{\beta}_{1jk}] = \beta_{1jk}$ and $\mathbb{V}ar(\hat{\beta}_{1jk}) = \displaystyle\frac{\sigma^2_{jk}}{n(1 - \rho_{jk})}$, we have  $\hat{\beta}_{1jk} \sim \mathcal{N}\left(\beta_{1jk}, \, \displaystyle\frac{\sigma^2_{jk}}{n(1 - \rho_{jk})}\right)$. Similarly to the formulation in step $1$, the estimate of the variance $\sigma^2$ is $\hat{\sigma}^2_{jk} =  \displaystyle\frac{1}{n - (k+1)} Y^\top P_{jk} Y$ where 
\begin{align*}
    P_{jk} & = I_n - \frac{a_n a_n^\top}{n} - \frac{X^{(k)} W_k^{-1} X^{(k)\top}}{n}  - \frac{X^{(k)} W_k^{-1} r_{jk} r_{jk}^\top W_k^{-1} X^{(k)\top}}{n(1 - \rho_{jk})} \\
    & - \frac{X_{*j} X_{*j}^\top}{n(1 - \rho_{jk})} + \frac{X^{(k)} W_k^{-1} r_{jk} X_{*j}^\top}{n(1 - \rho_{jk})} + \frac{X_{*j} r_{jk}^\top W_k^{-1} X^{(k)\top} }{n(1 - \rho_{jk})}
\end{align*}
  is a projection matrix.

\noindent Also, under $\mathbb{H}_0^{(jk)}$, $Y = a_n \beta_{0jk} + X^{(k)} \beta_{2jk} + \varepsilon_{jk}$, the OLS estimates are $\tilde{\beta}_{0jk} = n^{-1} a_n^\top Y$ and $\tilde{\beta}_{2jk} = n^{-1} W_k^{-1} X^{(k)\top}Y$, and with the idempotent projection $P_{jk}^0 = I_n - \frac{1}{n} a_n a_n^\top - \frac{1}{n} X^{(k)} W_k^{-1} X^{(k)\top}$ the null variance estimate is $\tilde{\sigma}^2_{jk} = Y^\top P_{jk}^0 Y/(n - k)$. The two test statistics can then be written as

\begin{eqnarray*}
    Z_{jk}^{(W)} &=& \frac{\hat{\beta}_{1jk}}{\sqrt{\widehat{\mathrm{Var}}(\hat\beta_{1jk})}}
    \;=\; \frac{(X_{*j}-X^{(k)}W_k^{-1}r_{jk})^\top Y}{\sqrt{n(1-\rho_{jk})\hat\sigma^2_{jk}}}
\end{eqnarray*}
and
$$Z_{jk}^{(S)} = \frac{\left(X_{*j} -  X^{(k)} W_k^{-1} r_{jk}\right)^{\top}Y}{\sqrt{n(1-\rho_{jk})\,\tilde{\sigma}^2_{jk}}}.$$

\subsection*{B2 : Distributions of $Z_{jk}^{(W)}$ and $Z_{jk}^{(S)}$}

\noindent The same argument as for step 1 gives $\sqrt{n (1- \rho_{jk})}\lambda_{jk}$ and $1 + \displaystyle \frac{(1- \rho_{jk})\lambda_{jk}^2}{2}$ as the mean and variance of $Z_{jk}^{(W)}$, respectively. Thus, $Z_{jk}^{(W)}$ is asymptotically normal : $Z_{jk}^{(W)}\sim \mathcal{N}\left(\sqrt{n (1- \rho_{jk})}\lambda_{jk}, \,  1 + \displaystyle \frac{(1- \rho_{jk})\lambda_{jk}^2}{2}\right)$.

\noindent As at step~1, write $Z_{jk}^{(S)}/\sqrt{n(1-\rho_{jk})} = g(\hat{s}, \hat{t})$ with $g(s,t) = s/\sqrt{t}$, $\hat{s} = (X_{*j} -  X^{(k)} W_k^{-1} r_{jk})^{\top}Y/\{n(1-\rho_{jk})\sigma_{jk}\}$ and $\hat{t} = Y^\top P_{jk}^0 Y/(n\sigma^2_{jk})$. Direct computation gives $\mathbb{E}[\hat{s}] = \lambda_{jk}$ and $\mathbb{V}ar(\hat{s}) = 1/\{n(1-\rho_{jk})\}$; by Corollary~2 of Theorem~5.5 in \citet{Chapter5Rencher}, $n\hat{t} \sim \chi^2\{n-k, n (1 - \rho_{jk})\lambda_{jk}^2/2\}$ in that reference's non-centrality convention, so that $\mathbb{E}[\hat{t}] \approx 1 + \lambda_{jk}^2(1 - \rho_{jk})$ and $\mathbb{V}ar(\hat{t}) \approx 2\{1 + 2\lambda_{jk}^2(1 - \rho_{jk})\}/n$; and $\mathbb{C}ov(\hat{s}, \hat{t}) = 2 \lambda_{jk}/n$. The first-order expansion of $g$ then gives
$$ \mathbb{E}[Z_{jk}^{(S)}] = \frac{\sqrt{n(1 - \rho_{jk})}\,\lambda_{jk}}{\sqrt{1 + (1 - \rho_{jk})\lambda_{jk}^2}}, \qquad
\mathbb{V}ar(Z_{jk}^{(S)}) = \displaystyle\frac{2 +(1 - \rho_{jk}) \lambda_{jk}^2}{2\left[1 + (1 - \rho_{jk})\lambda_{jk}^2\right]^3},$$
\noindent so that $Z_{jk}^{(S)}$ asymptotically follows the normal distribution with this mean and variance.

\noindent Here again, we recall that $\displaystyle\hat{\beta}_{1jk} = \frac{(X_{*j} - X^{(k)} W_k^{-1} r_{jk})^\top Y}{ n (1- \rho_{jk})} \sim \mathcal{N}\left(\beta_{1jk}, \frac{\sigma^2_{jk}}{n(1- \rho_{jk})}\right)$ and $\displaystyle\frac{(n-k-1) \hat{\sigma}^2_{jk}}{\sigma^2_{jk}} \sim \chi^2_{n-k-1}$, with $\hat{\beta}_{1jk}$ and $\hat{\sigma}^2_{jk}$ independent. It follows that $\sqrt{n (1- \rho_{jk})}\, \hat{\beta}_{1jk} \sim \mathcal{N}\!\left(\sqrt{n (1- \rho_{jk})}\,\beta_{1jk},\, \sigma^2_{jk}\right)$, so that using section 5.4.2 of \citet{Chapter5Rencher}, the Wald statistic follows a non-central Student distribution:
$$Z_{jk}^{(W)} = \frac{\sqrt{n (1- \rho_{jk})} \hat{\beta}_{1jk} /  \sigma_{jk}}{\sqrt{\frac{(n-k-1) \hat{\sigma}^2_{jk}}{\sigma^2_{jk}}/ (n-k-1)}} =   \frac{\sqrt{n(1- \rho_{jk})}  \hat{\beta}_{1jk}}{\sqrt{\hat{\sigma}^2_{jk}}}.$$

\noindent That is $Z_{jk}^{(W)}  \sim t\!\left(n-k-1,\; \lambda_{jk}\sqrt{n(1- \rho_{jk})}\right)$.

\subsection*{C1: Statistic and distribution of the extension}

\noindent From this section onward, the non-centrality of the chi-square and $F$ laws follows the convention of the \texttt{R} densities (\texttt{dchisq}, \texttt{df}), in which a non-central $\chi^2_{\nu}(\delta)$ has mean $\nu + \delta$; the parametrization of \citet{Chapter5Rencher} used in Sections A2 and B2 sets its non-centrality to half this quantity. The two conventions describe the same laws.

Under the model $Y = a_n \beta_{0jk} + \tilde X_{jk}\, \tilde\beta_{jk} + \varepsilon$, where $\tilde X_{jk} = \begin{pmatrix}
X_{*j} & X^{(k)}
\end{pmatrix}$ is a $n \times k$ matrix, $\tilde\beta_{jk} = \begin{pmatrix}
\beta_{1jk} & \beta_{2jk}^{\top}
\end{pmatrix}^{\top}$ is a regression coefficient vector of dimension $k$ and $\varepsilon \sim \mathcal{N}(0, \sigma_{jk}^2I_n)$. Here, the estimator $\hat{\tilde\beta}_{jk} =  [\tilde X_{jk}^{\top}  \tilde X_{jk}]^{-1}\tilde X_{jk}^{\top}Y$ follows a normal distribution, so that
\begin{align*}
\hat{\tilde\beta}_{jk} &\sim \mathcal{N}\left(\tilde\beta_{jk}, \, \sigma^2_{jk} [\tilde X_{jk}^{\top}  \tilde X_{jk}]^{-1}\right), \\
\frac{[\tilde X_{jk}^{\top}  \tilde X_{jk}]^{1/2}}{\sigma_{jk}}\, \hat{\tilde\beta}_{jk} &\sim \mathcal{N}\left(\frac{[\tilde X_{jk}^{\top}  \tilde X_{jk}]^{1/2}}{\sigma_{jk}}\, \tilde \beta_{jk}, \, I_k \right).
\end{align*}
Then, as described in section 5.3 of \citet{Chapter5Rencher}, the squared norm of this standardized vector satisfies $$ \frac{\hat{\tilde \beta}_{jk}^{\top} \tilde X_{jk}^{\top}  \tilde X_{jk} \hat{\tilde \beta}_{jk}}{\sigma_{jk}^2} = \frac{Y^\top \tilde X_{jk}(\tilde X_{jk}^\top \tilde X_{jk})^{-1}\tilde X_{jk}^\top Y}{\sigma^2_{jk}} \sim \chi^2\left(k, \tilde\delta_{jk}\right),$$
a non-central chi-square distribution with non-centrality parameter
$\tilde\delta_{jk} = \tilde\Lambda_{jk}^\top \tilde  X_{jk}^\top \tilde X_{jk}\, \tilde\Lambda_{jk},$ where $\tilde\Lambda_{jk}=\tilde\beta_{jk}/\sigma_{jk}=(\lambda_{jk},\lambda_{l_1},\lambda_{l_2},\cdots,\lambda_{l_{k-1}})^\top$.

\noindent Since $\displaystyle\frac{(n-k-1) \hat{\sigma}^2_{jk}}{\sigma^2_{jk}} \sim \chi^2_{n-k-1}$, following section 5.4.1 of \citet{Chapter5Rencher}, it is straightforward that $\tilde Z_{jk}^{(W)}$ follows a noncentral F distribution with $k$ and $n-k-1$ degrees of freedom and noncentrality parameter $\tilde\delta_{jk}$:

$$ \tilde Z_{jk}^{(W)} = \frac{\frac{\hat{\tilde \beta}_{jk}^{\top} \tilde X_{jk}^{\top}  \tilde X_{jk} \hat{\tilde\beta}_{jk}}{\sigma_{jk}^2} /k}{\frac{(n-k-1) \hat{\sigma}^2_{jk}}{\sigma^2_{jk}} /(n-k-1)} = \frac{\hat{\tilde\beta}_{jk}^{\top} \tilde X_{jk}^{\top}  \tilde X_{jk} \hat{\tilde \beta}_{jk}}{k \hat{\sigma}^2_{jk}} \sim F_{k, n-k-1}\left(\tilde\delta_{jk}\right).$$

\subsection*{C2: Exact prior integrals for the Multiple-SNP score}

\noindent This appendix explains how the prior integrals $\tilde A_{jk}$ and $\tilde B_{jk}$ of Section~2 are computed exactly. In words, averaging the effects over their Gaussian prior turns the non-central statistic into a weighted sum of chi-squared variables, whose density is a convergent mixture of central densities with explicit weights. Two classical facts are enough to show this, namely that a non-central chi-squared or Fisher variable is a Poisson mixture of central ones, and that the generating function of a Gaussian quadratic form has a closed form. We keep the notation of Section~2 and write $\nu = n-(k+1)$, $\mu_1,\dots,\mu_k\geq 0$ for the eigenvalues of $\tilde X_{jk}^\top\tilde X_{jk}$, $w_i = 1+\tau^2\mu_i$ and $\theta_i = \tau^2\mu_i/(1+\tau^2\mu_i) = 1-1/w_i$, for $i=1,\dots,k$.

\noindent \textbf{Mixture representation.} Conditionally on a Poisson variable $J$ with mean $\delta/2$, a non-central $\chi^2_k(\delta)$ variable is a central $\chi^2_{k+2J}$ variable \citep[chapters~29 and~30]{johnson1995}. Given $J=s$, $\tilde Z^{(W)}_{jk}$ is therefore distributed as $\frac{k+2s}{k}F_{k+2s,\nu}$, and the non-central Fisher density is the Poisson mixture
$$f_{k,\nu}(z;\delta) = \sum_{s\geq 0}e^{-\delta/2}\frac{(\delta/2)^s}{s!}\,h^{F}_s(z),\qquad h^{F}_s(z) = \frac{k}{k+2s}\,f_{k+2s,\nu}\Big(\frac{kz}{k+2s};0\Big).$$
Averaging over the prior distribution of $\tilde\delta_{jk}$, and exchanging the sum and the expectation, which is allowed because all the terms are positive, gives
$$\tilde A_{jk} = \sum_{s\geq 0}\omega_s\,h^{F}_s\big(\tilde z^{(W)}_{jk}\big),\qquad \omega_s = \mathbb E\left\{e^{-\tilde\delta_{jk}/2}\,\frac{(\tilde\delta_{jk}/2)^s}{s!}\right\}.$$
For NaM, the same argument on the chi-squared scale gives $\tilde A_{jk} = \sum_{s\geq0}\omega_s\,h^{\chi}_s\big(k\tilde z^{(W)}_{jk}\big)$, where $h^{\chi}_s(z) = f_{\chi^2_{k+2s}}(z)$ is the central chi-squared density with $k+2s$ degrees of freedom.

\noindent \textbf{The mixing weights.} The weights $\omega_s$ are nonnegative and sum to one, so they are the probabilities $P(N=s)$ of a count variable $N$, which we identify through its generating function. For $0\leq t\leq 1$,
$$\sum_{s\geq 0}\omega_s t^s = \mathbb E\left\{e^{-(1-t)\tilde\delta_{jk}/2}\right\}.$$
Under the prior $\tilde\Lambda_{jk}\sim\mathcal N(\mathbf 0,\tau^2 I_k)$, the non-centrality $\tilde\delta_{jk} = \tilde\Lambda_{jk}^\top\tilde X_{jk}^\top\tilde X_{jk}\tilde\Lambda_{jk}$ has the law of $\sum_{i=1}^k\tau^2\mu_iV_i^2$, where $V_1,\dots,V_k$ are independent standard normal variables, and $\mathbb E\{e^{-tV_i^2}\} = (1+2t)^{-1/2}$ for $t\geq 0$. Hence
$$\sum_{s\geq 0}\omega_s t^s = \prod_{i=1}^k\big\{1+(1-t)\tau^2\mu_i\big\}^{-1/2} = \prod_{i=1}^k\left(\frac{1-\theta_i}{1-\theta_i t}\right)^{1/2}.$$
Each factor on the right is the generating function of a negative binomial variable with shape $1/2$ and probability $\theta_i$, so the weights $\omega_s$ are the probabilities of $N = N_1+\dots+N_k$, a sum of $k$ independent such variables, as stated in Section~2. The prior integral $\tilde B_{jk}$ is given by the same series, with the same kernels and with weights computed from the $k-1$ eigenvalues of $X^{(k)\top}X^{(k)}$, since $\Lambda_k\sim\mathcal N(\mathbf 0,\tau^2I_{k-1})$.

\noindent \textbf{Computing the weights.} Write $\varphi(t) = \sum_{s\geq0}\omega_s t^s$ for the generating function above. Its logarithmic derivative is
$$\frac{\varphi'(t)}{\varphi(t)} = \frac12\sum_{i=1}^k\frac{\theta_i}{1-\theta_i t} = \frac12\sum_{u\geq 1}\vartheta_u\,t^{u-1},\qquad \vartheta_u = \sum_{i=1}^k\theta_i^{\,u}.$$
Multiplying by $\varphi(t)$ and identifying the coefficients of $t^{s-1}$ on both sides gives
$$\omega_0 = \prod_{i=1}^k(1-\theta_i)^{1/2},\qquad s\,\omega_s = \frac12\sum_{u=1}^{s}\vartheta_u\,\omega_{s-u}\quad(s\geq 1).$$
All the terms of this recursion are positive, so it involves no cancellation and is numerically stable.

\noindent \textbf{Truncation with a guaranteed error.} The series is stopped after $T$ terms, with a bound on the neglected part. The logarithm of the Fisher kernel is
$$\log h^{F}_s(z) = c+\log\Gamma\Big(\frac{k+2s+\nu}{2}\Big)-\log\Gamma\Big(\frac{k+2s}{2}\Big)+\frac{k+2s}{2}\log\frac{kz}{\nu+kz},$$
where $c$ does not depend on $s$. Its second derivative in $s$ equals $\psi'(\frac{k+2s+\nu}{2})-\psi'(\frac{k+2s}{2})$, which is negative because the trigamma function $\psi'$ is decreasing. Likewise,
$$\log h^{\chi}_s(z) = \Big(\frac{k+2s}{2}-1\Big)\log z-\frac z2-\frac{k+2s}{2}\log2-\log\Gamma\Big(\frac{k+2s}{2}\Big)$$
has second derivative $-\psi'(\frac{k+2s}{2})<0$. The kernels are therefore unimodal in $s$, first increasing and then decreasing. Once $T$ is beyond the mode, every neglected kernel is at most $h_{T+1}$, and the neglected part of the series satisfies
$$0\leq\tilde A_{jk}-\sum_{s=0}^{T}\omega_s\,h_s\leq h_{T+1}\,P(N>T)\leq h_{T+1}.$$
We increase $T$ until this bound is below $10^{-12}$ times the partial sum.

\noindent \textbf{Strong signals.} The kernels peak near $s\approx(k\tilde z^{(W)}_{jk}-k)/2$, so that a strong signal requires many terms. A rescaled form of the same series, due to \citet{ruben1962}, reduces this number (the rescaling constant is written $\kappa$ here, to keep $\beta$ for the regression coefficients). Under the prior, the numerator quadratic form of $\tilde Z^{(W)}_{jk}$ is a $\chi^2_{k+2N}$ variable, that is, it has the law of $\sum_iw_iV_i^2$. For any $0<\kappa\leq\min_iw_i$, this weighted sum also has the law of $\kappa\,\chi^2_{k+2N_\kappa}$, where $N_\kappa$ is defined as $N$ with $\theta_i$ replaced by $\theta_i(\kappa) = 1-\kappa/w_i$. This follows from the moment generating functions, since for $t<1/(2\max_iw_i)$
$$\mathbb E\big\{e^{t\kappa\chi^2_{k+2N_\kappa}}\big\} = (1-2t\kappa)^{-k/2}\prod_{i=1}^k\Big(\frac{\kappa}{w_i}\Big)^{1/2}\Big\{1-\frac{\theta_i(\kappa)}{1-2t\kappa}\Big\}^{-1/2} = \prod_{i=1}^k(1-2tw_i)^{-1/2}.$$
The representation above is the case $\kappa=1$. With $\kappa=\min_iw_i$, one of the $\theta_i(\kappa)$ vanishes and the others decrease, the kernels become $\kappa^{-1}h_s(\cdot/\kappa)$, and their mode moves to about $(k\tilde z^{(W)}_{jk}/\kappa-k)/2$, so that the number of terms is divided by about $\kappa$. For $\tilde B_{jk}$, the weight of the candidate coordinate equals one, so that $\kappa = 1$; these weights are shared by all the candidates of a step and are computed once.


\subsection*{Single-locus simulation results}\label{subsec:supp_locus_sim}

Tables~\ref{tab:locus_sim_informative_fixed} and~\ref{tab:locus_sim_informative_adaptive} report the single-locus comparison under the informative prior.

\begin{table}[htbp]
\centering
\caption{\textbf{Single-locus fixed-$K$ comparison.} Performance under the informative prior with $m_{\mathrm{locus}}=200$, $\rho=0.9$ and $K=3$.}
\label{tab:locus_sim_informative_fixed}
\setlength{\tabcolsep}{3.5pt}
\renewcommand{\arraystretch}{1.0}
\scriptsize
\begin{tabular}{c l c c c c c c}
\toprule
$n$ & \textbf{Method} & $\hat M$ & $\hat K$ & Precision & Recall & F1 & FDR \\
\midrule
\multirow{13}{*}{$1000$} & NaW & 2.40 & 3.0 & 0.800 & 0.800 & 0.800 & 0.200 \\
 & AsW & 2.46 & 3.0 & 0.822 & 0.822 & 0.822 & 0.178 \\
 & ExW & 2.50 & 3.0 & 0.835 & 0.835 & 0.835 & 0.165 \\
 & AsS & 2.48 & 3.0 & 0.827 & 0.827 & 0.827 & 0.173 \\
 & NaM & \textbf{2.87} & \textbf{3.0} & \textbf{0.955} & \textbf{0.955} & \textbf{0.955} & \textbf{0.045} \\
 & ExM & 2.87 & 3.0 & 0.955 & 0.955 & 0.955 & 0.045 \\
 & SuSiE\textsubscript{unif} & 0.83 & 3.0 & 0.278 & 0.278 & 0.278 & 0.722 \\
 & SuSiE\textsubscript{prior} & 2.31 & 3.0 & 0.768 & 0.768 & 0.768 & 0.232 \\
 & varbvs & 0.96 & 3.0 & 0.322 & 0.322 & 0.322 & 0.678 \\
 & LASSO\textsubscript{topK} & 0.00 & 0.0 & {NA} & 0.000 & 0.000 & {NA} \\
 & LASSO\textsubscript{1se} & 0.00 & 0.0 & {NA} & 0.000 & 0.000 & {NA} \\
 & CARMA & 0.94 & 3.0 & 0.312 & 0.312 & 0.312 & 0.688 \\
 & FINEMAP & 0.94 & 3.0 & 0.312 & 0.312 & 0.312 & 0.688 \\
\bottomrule
\end{tabular}
\end{table}

\begin{table}[htbp]
\centering
\caption{\textbf{Single-locus adaptive comparison.} Performance under the informative prior using each method's stated stopping rule.}
\label{tab:locus_sim_informative_adaptive}
\setlength{\tabcolsep}{3.5pt}
\renewcommand{\arraystretch}{1.0}
\scriptsize
\begin{tabular}{c l l c c c c c c}
\toprule
$n$ & \textbf{Method} & \textbf{Criterion} & $\hat M$ & $\hat K$ & Precision & Recall & F1 & FDR \\
\midrule
\multirow{48}{*}{$1000$} & \multirow{6}{*}{NaW} & $R^2$ & 2.58 & 19.9 & 0.270 & 0.862 & 0.322 & 0.730 \\
 &  & $BIC$ & 1.24 & 1.8 & 0.757 & 0.413 & 0.502 & 0.243 \\
 &  & $eBIC$ & 0.87 & 1.0 & 0.865 & 0.290 & 0.434 & 0.135 \\
 &  & $Post(0.5)$ & 2.77 & 4.8 & 0.624 & 0.925 & 0.726 & 0.376 \\
 &  & $Post(0.95)$ & 1.04 & 1.1 & 0.898 & 0.347 & 0.441 & 0.102 \\
 &  & $Post(0.99)$ & 0.52 & 0.5 & 0.971 & 0.173 & 0.239 & 0.029 \\
\cmidrule(lr){2-9}
 & \multirow{6}{*}{AsW} & $R^2$ & 2.65 & 21.6 & 0.206 & 0.883 & 0.279 & 0.794 \\
 &  & $BIC$ & 1.31 & 1.7 & 0.847 & 0.438 & 0.545 & 0.153 \\
 &  & $eBIC$ & 0.95 & 1.0 & 0.945 & 0.317 & 0.474 & 0.055 \\
 &  & $Post(0.5)$ & 2.71 & 5.0 & 0.594 & 0.905 & 0.699 & 0.406 \\
 &  & $Post(0.95)$ & 1.10 & 1.1 & 0.962 & 0.367 & 0.469 & 0.038 \\
 &  & $Post(0.99)$ & 0.39 & 0.4 & 0.964 & 0.128 & 0.183 & 0.036 \\
\cmidrule(lr){2-9}
 & \multirow{6}{*}{ExW} & $R^2$ & 2.64 & 21.2 & 0.222 & 0.880 & 0.289 & 0.778 \\
 &  & $BIC$ & 1.31 & 1.7 & 0.860 & 0.438 & 0.550 & 0.140 \\
 &  & $eBIC$ & 0.95 & 1.0 & 0.950 & 0.318 & 0.477 & 0.050 \\
 &  & $Post(0.5)$ & 2.66 & 4.3 & 0.668 & 0.887 & 0.743 & 0.332 \\
 &  & $Post(0.95)$ & 1.09 & 1.1 & 0.962 & 0.363 & 0.465 & 0.038 \\
 &  & $Post(0.99)$ & 0.38 & 0.4 & 0.963 & 0.125 & 0.178 & 0.037 \\
\cmidrule(lr){2-9}
 & \multirow{6}{*}{AsS} & $R^2$ & 2.65 & 21.6 & 0.207 & 0.883 & 0.280 & 0.793 \\
 &  & $BIC$ & 1.31 & 1.7 & 0.854 & 0.438 & 0.547 & 0.146 \\
 &  & $eBIC$ & 0.95 & 1.0 & 0.950 & 0.318 & 0.477 & 0.050 \\
 &  & $Post(0.5)$ & 2.71 & 4.9 & 0.600 & 0.905 & 0.702 & 0.400 \\
 &  & $Post(0.95)$ & 1.09 & 1.1 & 0.962 & 0.363 & 0.465 & 0.038 \\
 &  & $Post(0.99)$ & 0.35 & 0.4 & 0.962 & 0.118 & 0.169 & 0.038 \\
\cmidrule(lr){2-9}
 & \multirow{6}{*}{NaM} & $R^2$ & 2.62 & 3.5 & 0.780 & 0.873 & 0.799 & 0.220 \\
 &  & $BIC$ & 1.25 & 1.4 & 0.880 & 0.415 & 0.543 & 0.120 \\
 &  & $eBIC$ & 0.87 & 1.0 & 0.865 & 0.290 & 0.434 & 0.135 \\
 &  & $Post(0.5)$ & \textbf{3.00} & \textbf{3.1} & \textbf{0.966} & \textbf{1.000} & \textbf{0.981} & \textbf{0.034} \\
 &  & $Post(0.95)$ & 0.67 & 0.7 & 0.899 & 0.223 & 0.335 & 0.101 \\
 &  & $Post(0.99)$ & 0.41 & 0.4 & 0.965 & 0.138 & 0.207 & 0.035 \\
\cmidrule(lr){2-9}
 & \multirow{6}{*}{ExM} & $R^2$ & 2.62 & 3.5 & 0.780 & 0.873 & 0.799 & 0.220 \\
 &  & $BIC$ & 1.25 & 1.4 & 0.880 & 0.415 & 0.543 & 0.120 \\
 &  & $eBIC$ & 0.87 & 1.0 & 0.865 & 0.290 & 0.434 & 0.135 \\
 &  & $Post(0.5)$ & 3.00 & 3.1 & 0.966 & 1.000 & 0.981 & 0.034 \\
 &  & $Post(0.95)$ & 0.67 & 0.7 & 0.899 & 0.223 & 0.335 & 0.101 \\
 &  & $Post(0.99)$ & 0.41 & 0.4 & 0.965 & 0.138 & 0.207 & 0.035 \\
\cmidrule(lr){2-9}
 & SuSiE\textsubscript{unif} & $Adaptive$ & 0.02 & 0.0 & 0.875 & 0.007 & 0.010 & 0.125 \\
 & SuSiE\textsubscript{prior} & $Adaptive$ & 0.07 & 0.1 & 0.929 & 0.023 & 0.034 & 0.071 \\
 & varbvs & $PIP{>}0.5$ & 0.48 & 1.0 & 0.502 & 0.162 & 0.207 & 0.498 \\
 & varbvs & $PIP{>}0.95$ & 0.12 & 0.2 & 0.590 & 0.040 & 0.056 & 0.410 \\
 & varbvs & $PIP{>}0.99$ & 0.06 & 0.1 & 0.750 & 0.020 & 0.029 & 0.250 \\
 & LASSO\textsubscript{1se} & $Adaptive$ & 0.01 & 0.0 & 1.000 & 0.002 & 0.002 & 0.000 \\
 & CARMA & $PIP{>}0.5$ & 0.11 & 0.2 & 0.661 & 0.037 & 0.053 & 0.339 \\
 & CARMA & $PIP{>}0.95$ & 0.01 & 0.0 & 1.000 & 0.005 & 0.007 & 0.000 \\
 & CARMA & $PIP{>}0.99$ & 0.01 & 0.0 & 1.000 & 0.003 & 0.005 & 0.000 \\
 & FINEMAP & $PIP{>}0.5$ & 0.10 & 0.2 & 0.679 & 0.035 & 0.051 & 0.321 \\
 & FINEMAP & $PIP{>}0.95$ & 0.01 & 0.0 & 1.000 & 0.005 & 0.007 & 0.000 \\
 & FINEMAP & $PIP{>}0.99$ & 0.01 & 0.0 & 1.000 & 0.003 & 0.005 & 0.000 \\
\bottomrule
\end{tabular}
\end{table}

Table~\ref{tab:carma_prior_sensitivity} adds a prior-informed CARMA variant, and Table~\ref{tab:locus_scenarios} extends the comparison to the four prior scenarios.

\begin{table}[htbp]
\centering
\caption{\textbf{Prior-informed CARMA sensitivity.} F1 in the single-locus design under informative and uniform priors.}
\label{tab:carma_prior_sensitivity}
\small
\setlength{\tabcolsep}{6pt}
\renewcommand{\arraystretch}{1.12}
\begin{tabular}{llcccc}
\toprule
\textbf{Scenario} & \textbf{Regime} & \textbf{ExM} & \textbf{SuSiE$_p$} & \textbf{CARMA} & \textbf{CARMA$_p$} \\
\midrule
Informative & Fixed    & 0.96 & 0.77 & 0.31 & 0.65 \\
Informative & Adaptive & 0.98 & 0.03 & 0.05 & 0.42 \\
Uniform     & Fixed    & 0.27 & 0.28 & 0.31 & 0.32 \\
Uniform     & Adaptive & 0.29 & 0.01 & 0.05 & 0.05 \\
\bottomrule
\end{tabular}
\end{table}

\begin{table}[htbp]
\centering
\caption{\textbf{Single-locus prior sensitivity.} Mean causal variants recovered out of $3$ at fixed $K=3$ across the four prior scenarios.}
\label{tab:locus_scenarios}
\small
\setlength{\tabcolsep}{6pt}
\begin{tabular}{lcccc}
\toprule
Method & Informative & Uniform & Misspecified & Mixed \\
\midrule
NaW & 2.40 & 0.90 & 0.42 & 0.96 \\
AsW & 2.47 & 0.90 & 0.33 & 0.96 \\
ExW & 2.51 & 0.92 & 0.32 & 0.98 \\
AsS & 2.48 & 0.90 & 0.32 & 0.96 \\
NaM & 2.87 & 0.81 & 0.21 & 1.25 \\
ExM & 2.87 & 0.81 & 0.21 & 1.25 \\
SuSiE\textsubscript{unif} & 0.84 & 0.84 & 0.84 & 0.84 \\
SuSiE\textsubscript{prior} & 2.31 & 0.84 & 0.22 & 0.86 \\
varbvs & 0.97 & 0.87 & 0.88 & 0.87 \\
CARMA & 0.94 & 0.93 & 0.93 & 0.92 \\
CARMA\textsubscript{p} & 1.94 & 0.95 & 2.09 & 0.96 \\
FINEMAP & 0.94 & 0.94 & 0.94 & 0.94 \\
LASSO\textsubscript{1se} & 0.00 & 0.01 & 0.01 & 0.01 \\
\bottomrule
\end{tabular}
\end{table}

\subsection*{Controlled AR(1) sensitivity analyses}\label{subsec:supp_scenarios}

Table~\ref{tab:synth_informative_fixed} gives the oracle-prior fixed-$K$ reference, and Table~\ref{tab:synth_scenarios} repeats the evaluation across the prior scenarios of Table~\ref{tab:parameters} and both sample sizes.
\begin{table}[htbp]
\centering
\caption{\textbf{Oracle-prior AR(1) sensitivity.} F1 at fixed $K=15$ across the three LD levels for $n=1000$. The prior-only row ranks by $p_j$ with marginal $|r|$ used to break ties.}
\label{tab:synth_informative_fixed}
\setlength{\tabcolsep}{4pt}
\renewcommand{\arraystretch}{1.0}
\small
\begin{tabular}{lccc}
\toprule
\textbf{Method} & \multicolumn{3}{c}{$n = 1000$} \\
\cmidrule(lr){2-4}
 & $\rho=0.80$ & $\rho=0.95$ & $\rho=0.98$ \\
\midrule
Prior-only ranking & 1.000 & 1.000 & 1.000 \\
\midrule
NaW & 0.480 & 0.487 & 0.479 \\
AsW & 0.588 & 0.608 & 0.604 \\
ExW & 0.599 & 0.613 & 0.623 \\
AsS & 0.593 & 0.612 & 0.612 \\
NaM & 0.992 & 0.993 & 0.984 \\
ExM & 0.993 & 0.993 & 0.984 \\
SuSiE\textsubscript{unif} & 0.137 & 0.143 & 0.119 \\
SuSiE\textsubscript{prior} & 0.452 & 0.485 & 0.461 \\
LASSO\textsubscript{topK} & 0.000 & 0.000 & 0.008 \\
LASSO\textsubscript{1se} & 0.000 & 0.000 & 0.003 \\
BayesC & 0.132 & 0.096 & 0.071 \\
\bottomrule
\end{tabular}
\end{table}

\begin{table}[htbp]
\centering
\caption{\textbf{Controlled AR(1) results across prior scenarios.} Mean causal variants recovered out of $15$ at fixed $K=15$ for $n=80$ and $n=1000$, averaged over the three LD levels.}
\label{tab:synth_scenarios}
\scriptsize
\setlength{\tabcolsep}{4pt}
\begin{tabular}{lcccccccc}
\toprule
 & \multicolumn{4}{c}{$n=80$} & \multicolumn{4}{c}{$n=1000$} \\
\cmidrule(lr){2-5}\cmidrule(lr){6-9}
Method & Inf. & Unif. & Missp. & Mixed & Inf. & Unif. & Missp. & Mixed \\
\midrule
Prior-only ranking & 15.00 & 0.99 & 0.00 & 5.25 & 15.00 & 2.15 & 0.00 & 5.64 \\
\midrule
NaW & 4.35 & 0.38 & 0.11 & 0.60 & 7.23 & 1.91 & 0.79 & 2.41 \\
AsW & 4.31 & 0.38 & 0.11 & 0.60 & 9.00 & 1.92 & 0.74 & 2.70 \\
ExW & 6.94 & 0.36 & 0.07 & 0.65 & 9.17 & 1.92 & 0.70 & 2.75 \\
AsS & 6.93 & 0.39 & 0.04 & 0.73 & 9.09 & 1.92 & 0.73 & 2.73 \\
NaM & 14.66 & 0.40 & 0.06 & 5.15 & 14.85 & 1.19 & 0.13 & 5.19 \\
ExM & 14.81 & 0.45 & 0.05 & 5.14 & 14.85 & 1.18 & 0.13 & 5.19 \\
SuSiE\textsubscript{unif} & 0.93 & 0.93 & 0.93 & 0.93 & 1.99 & 1.99 & 1.99 & 1.99 \\
SuSiE\textsubscript{prior} & 8.35 & 0.93 & 0.18 & 1.88 & 6.99 & 1.99 & 0.77 & 2.21 \\
LASSO\textsubscript{topK} & 0.12 & 0.08 & 0.07 & 0.09 & 0.03 & 0.01 & 0.03 & 0.04 \\
LASSO\textsubscript{1se} & 0.09 & 0.09 & 0.09 & 0.09 & 0.03 & 0.06 & 0.06 & 0.04 \\
\bottomrule
\end{tabular}
\end{table}

Tables~\ref{tab:stop_joint}, \ref{tab:synth_informative_adaptive} and~\ref{tab:synth_adaptive_best} collect the stopping-rule sensitivities, and Table~\ref{tab:smalln_adaptive} gives the adaptive small-sample analogue.

\begin{table}[htbp]
\centering
\caption{\textbf{Stopping rules for ExM.} Mean causal variants recovered with mean selected size in brackets across controlled and single-locus designs.}
\label{tab:stop_joint}
\scriptsize
\setlength{\tabcolsep}{3pt}
\begin{tabular}{lcccccc}
\toprule
Design and prior & adj.~$R^2$ & BIC & eBIC & $q>0.5$ & $q>0.95$ & $q>0.99$ \\
\midrule
\multicolumn{7}{l}{\textit{AR(1), $m=5{,}000$}} \\
\quad Informative & 12.49 [12.7] & 4.65 [4.8] & 0.86 [1.0] & 15.00 [15.1] & 0.85 [1.0] & 0.79 [0.9] \\
\quad Uniform & 1.00 [4.6] & 0.96 [3.7] & 0.36 [1.0] & 0.59 [1.9] & 0.35 [1.0] & 0.35 [0.9] \\
\quad Misspecified & 0.13 [4.4] & 0.13 [3.5] & 0.13 [1.0] & 0.13 [1.8] & 0.13 [1.0] & 0.12 [0.9] \\
\quad Mixed & 4.39 [5.5] & 1.65 [2.2] & 0.43 [1.0] & 5.16 [5.7] & 0.43 [1.0] & 0.42 [0.9] \\
\quad Informative, $n=80$ & 11.71 [12.1] & 5.62 [5.8] & 0.81 [1.0] & 15.00 [15.2] & 0.80 [1.0] & 0.53 [0.6] \\
\addlinespace[2pt]
\multicolumn{7}{l}{\textit{Locus, $m_{\text{locus}}=200$}} \\
\quad Informative & 2.62 [3.5] & 1.25 [1.4] & 0.87 [1.0] & 3.00 [3.1] & 0.67 [0.7] & 0.42 [0.4] \\
\quad Uniform & 0.80 [2.8] & 0.74 [1.9] & 0.47 [1.0] & 0.53 [1.1] & 0.33 [0.6] & 0.20 [0.3] \\
\quad Misspecified & 0.21 [1.9] & 0.21 [1.0] & 0.21 [1.0] & 0.21 [4.0] & 0.14 [0.5] & 0.07 [0.2] \\
\quad Mixed & 1.10 [2.8] & 0.63 [1.4] & 0.50 [1.0] & 1.17 [1.7] & 0.41 [0.7] & 0.24 [0.4] \\
\bottomrule
\end{tabular}
\end{table}

\begin{table}[htbp]
\caption{\textbf{BIC-stopped AR(1) selection.} F1 with mean selected size in parentheses across the three LD levels for $n=1000$.}
\label{tab:synth_informative_adaptive}
\centering\small
\setlength{\tabcolsep}{4pt}
\begin{tabular}{lccc}
\toprule
\textbf{Method} & $\rho=0.80$ & $\rho=0.95$ & $\rho=0.98$ \\
\midrule
NaW & 0.391 (11.0) & 0.396 (9.7) & 0.378 (9.1) \\
AsW & 0.402 (9.8) & 0.409 (8.8) & 0.384 (8.3) \\
ExW & 0.412 (9.2) & 0.415 (8.1) & 0.397 (7.7) \\
AsS & 0.406 (9.8) & 0.411 (8.4) & 0.395 (7.9) \\
NaM & 0.483 (5.1) & 0.501 (5.3) & 0.466 (5.1) \\
ExM & 0.486 (5.1) & 0.501 (5.3) & 0.467 (5.1) \\
\bottomrule
\end{tabular}
\end{table}

\begin{table}[htbp]
\caption{\textbf{Adaptive stopping sensitivity.} Best F1 obtained by each method over the evaluated stopping rules, with the selected rule and mean size in brackets.}
\label{tab:synth_adaptive_best}
\centering\scriptsize
\setlength{\tabcolsep}{2.5pt}
\begin{tabular}{lccc}
\toprule
\textbf{Method} & $\rho=0.80$ & $\rho=0.95$ & $\rho=0.98$ \\
\midrule
NaW & 0.402 [$R^2$, 29.0] & 0.414 [$R^2$, 29.0] & 0.430 [$R^2$, 29.0] \\
AsW & 0.441 [$R^2$, 29.0] & 0.460 [$R^2$, 29.0] & 0.470 [$R^2$, 29.0] \\
ExW & 0.455 [$R^2$, 29.0] & 0.486 [$R^2$, 29.0] & 0.503 [$R^2$, 29.0] \\
AsS & 0.448 [$R^2$, 29.0] & 0.467 [$R^2$, 29.0] & 0.476 [$R^2$, 29.0] \\
NaM & 0.995 [$q>.5$, 15.1] & 0.996 [$q>.5$, 15.1] & 0.992 [$q>.5$, 15.2] \\
ExM & 0.996 [$q>.5$, 15.1] & 0.996 [$q>.5$, 15.1] & 0.992 [$q>.5$, 15.2] \\
SuSiE\textsubscript{unif} & 0.002 [native, 0.0] & 0.003 [native, 0.1] & 0.003 [native, 0.1] \\
SuSiE\textsubscript{prior} & 0.006 [native, 0.1] & 0.006 [native, 0.1] & 0.007 [native, 0.1] \\
LASSO & 0.008 [native, 0.9] & 0.002 [native, 0.0] & 0.003 [native, 0.2] \\
BayesC & 0.037 [PIP$>$.5, 193.3] & 0.007 [PIP$>$.5, 87.3] & 0.004 [PIP$>$.5, 78.4] \\
\bottomrule
\end{tabular}
\end{table}

\begin{table}[H]
\centering
\caption{\textbf{Small-sample adaptive results at $n=80$.} Best F1 with the selected rule and mean selected size for each in-house method.}
\label{tab:smalln_adaptive}
\scriptsize
\setlength{\tabcolsep}{3pt}
\renewcommand{\arraystretch}{1.12}
\begin{tabularx}{\textwidth}{lXXX}
\toprule
\textbf{Method} & $\gamma=0.2$ & $\gamma=0.6$ & $\gamma=0.9$ \\
\midrule
AsW & 0.177 [adj.~$R^2$, $\hat K$=29.0] & 0.257 [adj.~$R^2$, $\hat K$=29.0] & 0.352 [$q>0.95$, $\hat K$=9.1] \\
ExW & 0.537 [$q>0.5$, $\hat K$=30.0] & 0.589 [$q>0.5$, $\hat K$=29.9] & 0.623 [$q>0.5$, $\hat K$=30.0] \\
$\Delta$F1(ExW - AsW) & +0.360 & +0.332 & +0.270 \\
\addlinespace[2pt]
AsS & 0.374 [$q>0.5$, $\hat K$=30.0] & 0.439 [adj.~$R^2$, $\hat K$=28.6] & 0.529 [BIC, $\hat K$=20.4] \\
\addlinespace[2pt]
NaM & 0.981 [$q>0.5$, $\hat K$=15.6] & 0.989 [$q>0.5$, $\hat K$=15.3] & 0.985 [$q>0.5$, $\hat K$=15.5] \\
ExM & 0.986 [$q>0.5$, $\hat K$=15.4] & 0.994 [$q>0.5$, $\hat K$=15.2] & 0.997 [$q>0.5$, $\hat K$=15.1] \\
$\Delta$F1(ExM - NaM) & +0.004 & +0.005 & +0.012 \\
\bottomrule
\end{tabularx}
\end{table}

Table~\ref{tab:tau_sensitivity} varies the slab scale at $n=80$.

\begin{table}[htbp]
\caption{\textbf{Sensitivity to the slab scale.} Mean causal variants recovered out of $15$ at fixed $K=15$ for $\tau\in\{0.2,0.5,1\}$.}
\label{tab:tau_sensitivity}
\centering\small
\begin{tabular}{lccc}
\toprule
Method & $\tau=0.2$ & $\tau=0.5$ & $\tau=1$ \\
\midrule
NaW & 4.35 & 2.88 & 2.64 \\
AsW & 4.31 & 2.72 & 3.77 \\
ExW & 6.94 & 3.79 & 4.81 \\
AsS & 6.93 & 3.84 & 5.57 \\
NaM & 14.66 & 14.57 & 14.56 \\
ExM & 14.81 & 14.64 & 14.63 \\
\bottomrule
\end{tabular}
\end{table}

\subsection*{Numerical validation and finite-sample calibration}\label{subsec:supp_checks}

Table~\ref{tab:exact_valid} checks the deterministic Multiple-SNP integration against characteristic-function inversion for ExM and the Davies algorithm for NaM.

\begin{table}[htbp]
\centering
\caption{\textbf{Validation of the exact Multiple-SNP integral.} Relative numerical differences and Monte-Carlo error across sixteen configurations.}
\label{tab:exact_valid}
\scriptsize
\setlength{\tabcolsep}{2.5pt}
\begin{tabular}{rrcrrcccccc}
\toprule
$k$ & $n$ & $\tau$ & $n\tau^2$ & $T$ & $\tilde A_{jk}$ (ExM) & inversion (ExM) & Davies (NaM) & MC, $10^6$ & MC rel. s.e., $S=100$ \\
\midrule
1 & 80 & $0.2$ & $3.2$ & 64 & $0.0622$ & $0$ & $4.4\times 10^{-16}$ & $-5.6\times 10^{-4}$ & $0.0425$ \\
3 & 80 & $0.2$ & $3.2$ & 64 & $0.156$ & $0$ & $5.9\times 10^{-10}$ & $9.4\times 10^{-4}$ & $0.0463$ \\
8 & 80 & $0.2$ & $3.2$ & 128 & $0.0448$ & $6.7\times 10^{-16}$ & $2.1\times 10^{-8}$ & $-1.2\times 10^{-3}$ & $0.119$ \\
15 & 80 & $0.2$ & $3.2$ & 256 & $0.0401$ & $2.2\times 10^{-16}$ & $3.7\times 10^{-8}$ & $-9.1\times 10^{-4}$ & $0.142$ \\
1 & 1000 & $0.2$ & $40.0$ & 64 & $0.0221$ & $1.3\times 10^{-15}$ & $1.1\times 10^{-16}$ & $6.1\times 10^{-5}$ & $0.121$ \\
3 & 1000 & $0.2$ & $40.0$ & 64 & $0.0117$ & $1.9\times 10^{-11}$ & $5.6\times 10^{-10}$ & $-2.9\times 10^{-3}$ & $0.385$ \\
8 & 1000 & $0.2$ & $40.0$ & 128 & $6.28\times 10^{-3}$ & $4.1\times 10^{-9}$ & $9.4\times 10^{-10}$ & $-2.6\times 10^{-3}$ & $0.460$ \\
15 & 1000 & $0.2$ & $40.0$ & 256 & $4.14\times 10^{-3}$ & $4.4\times 10^{-11}$ & $9.1\times 10^{-9}$ & $-0.010$ & $0.597$ \\
\midrule
1 & 80 & $0.5$ & $20.0$ & 64 & $0.0135$ & $8.9\times 10^{-16}$ & $4.4\times 10^{-16}$ & $-1.5\times 10^{-3}$ & $0.116$ \\
3 & 80 & $0.5$ & $20.0$ & 64 & $0.0289$ & $8.9\times 10^{-16}$ & $6.5\times 10^{-10}$ & $-3.4\times 10^{-4}$ & $0.217$ \\
8 & 80 & $0.5$ & $20.0$ & 128 & $0.0169$ & $6.7\times 10^{-16}$ & $9.9\times 10^{-11}$ & $3.9\times 10^{-3}$ & $0.275$ \\
15 & 80 & $0.5$ & $20.0$ & 256 & $0.0151$ & $1.1\times 10^{-15}$ & $4.1\times 10^{-9}$ & $-4.3\times 10^{-3}$ & $0.297$ \\
1 & 1000 & $0.5$ & $250.0$ & 64 & $4.99\times 10^{-3}$ & $1.1\times 10^{-13}$ & $4.4\times 10^{-16}$ & $-2.5\times 10^{-3}$ & $0.220$ \\
3 & 1000 & $0.5$ & $250.0$ & 64 & $1.04\times 10^{-3}$ & $1.3\times 10^{-11}$ & $8.2\times 10^{-10}$ & $-6.2\times 10^{-3}$ & $1.08$ \\
8 & 1000 & $0.5$ & $250.0$ & 64 & $1.08\times 10^{-6}$ & $8.7\times 10^{-7}$ & $3.0\times 10^{-6}$ & $0.18$ & $31.08$ \\
15 & 1000 & $0.5$ & $250.0$ & 128 & $2.99\times 10^{-10}$ & $5.4\times 10^{-8}$ & $9.6\times 10^{-3}$ & $-1.00$ & $83.12$ \\
\bottomrule
\end{tabular}
\end{table}

The Monte-Carlo average loses accuracy once the prior spread of the non-centrality exceeds the observed signal, and Table~\ref{tab:exact_select} shows the consequence on selection.

\begin{table}[htbp]
\centering
\caption{\textbf{Exact and Monte-Carlo Multiple-SNP scores.} Mean causal variants recovered out of $15$ with F1 in brackets across values of $\tau$.}
\label{tab:exact_select}
\small
\begin{tabular}{lccccc}
\toprule
 & & \multicolumn{2}{c}{ExM} & \multicolumn{2}{c}{NaM} \\
\cmidrule(lr){3-4}\cmidrule(lr){5-6}
$\tau$ & AsW & MC, $S=30$ & exact & MC, $S=30$ & exact \\
\midrule
$0.070$ & $11.95$ [$0.80$] & $14.85$ [$0.99$] & $14.95$ [$1.00$] & $14.95$ [$1.00$] & $14.95$ [$1.00$] \\
$0.100$ & $11.40$ [$0.76$] & $14.60$ [$0.97$] & $15.00$ [$1.00$] & $14.65$ [$0.98$] & $15.00$ [$1.00$] \\
$0.141$ & $11.20$ [$0.75$] & $8.20$ [$0.55$] & $14.95$ [$1.00$] & $8.20$ [$0.55$] & $14.95$ [$1.00$] \\
$0.200$ & $11.45$ [$0.76$] & $4.30$ [$0.29$] & $14.95$ [$1.00$] & $4.55$ [$0.30$] & $14.95$ [$1.00$] \\
\bottomrule
\end{tabular}
\smallskip
{\footnotesize At $\tau=0.2$, increasing the Monte-Carlo draws from $300$ to $3000$ raises ExM recovery from $9.45$ to $13.65$ and NaM recovery from $9.50$ to $13.45$.}
\end{table}

Table~\ref{tab:greedy_check} compares the greedy NaM and ExM with the non-greedy SuSiE on a single high-LD block.

\begin{table}[htbp]
\centering
\caption{\textbf{Greedy selection under strong LD.} Mean causal LD clusters recovered out of $3$ under the informative prior.}
\label{tab:greedy_check}
\small
\begin{tabular}{lccc}
\toprule
$\rho$ & NaM & ExM & SuSiE \\
\midrule
$0.90$ & $3.00$ & $3.00$ & $3.00$ \\
$0.95$ & $2.62$ & $2.66$ & $2.98$ \\
\bottomrule
\end{tabular}
\end{table}

Table~\ref{tab:null_calibration} evaluates the exact conditional law under the global null, for a fixed reference candidate and for the selected one.

\begin{table}[htbp]
\caption{\textbf{Conditional null calibration.} Empirical frequency of $p<0.05$ for the selected candidate and for a fixed reference candidate under exact $t$ and normal reference laws.}
\label{tab:null_calibration}
\centering\scriptsize
\setlength{\tabcolsep}{3.5pt}
\begin{tabular}{lccccccccc}
\toprule
 & \multicolumn{3}{c}{$n=50$} & \multicolumn{3}{c}{$n=80$} & \multicolumn{3}{c}{$n=1000$} \\
\cmidrule(lr){2-4}\cmidrule(lr){5-7}\cmidrule(lr){8-10}
Step $k$ & sel. & ref.\ $t$ & ref.\ $\mathcal{N}$ & sel. & ref.\ $t$ & ref.\ $\mathcal{N}$ & sel. & ref.\ $t$ & ref.\ $\mathcal{N}$ \\
\midrule
1 & 1.000 & 0.040 & 0.050 & 1.000 & 0.056 & 0.058 & 1.000 & 0.060 & 0.060 \\
2 & 1.000 & 0.054 & 0.060 & 1.000 & 0.050 & 0.050 & 1.000 & 0.050 & 0.050 \\
5 & 1.000 & 0.048 & 0.052 & 1.000 & 0.042 & 0.052 & 1.000 & 0.036 & 0.036 \\
10 & 1.000 & 0.044 & 0.048 & 1.000 & 0.036 & 0.038 & 1.000 & 0.034 & 0.034 \\
\bottomrule
\end{tabular}
\end{table}

\begin{figure}[htbp]
\centering
\includegraphics[width=0.78\textwidth]{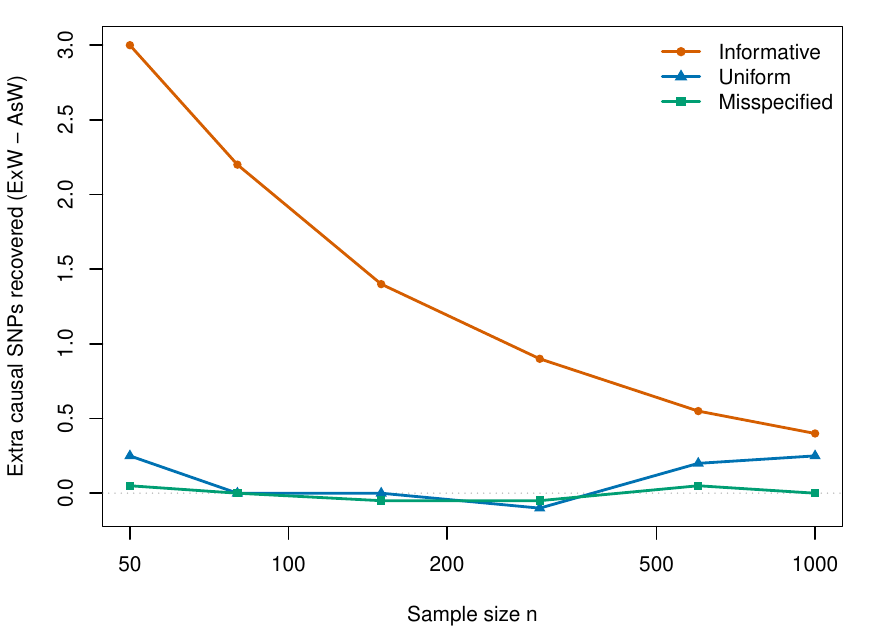}
\caption{\textbf{Exact-Wald gain by prior quality.} ExW minus AsW causal recovery across sample size under informative, uniform and misspecified priors.}
\label{fig:premium_by_prior}
\end{figure}

Figure~\ref{fig:premium_by_prior} repeats the sample-size sweep of the main text under the uniform and misspecified priors.

\subsection*{Computational scaling}\label{subsec:supp_scaling}

\begin{table}[htbp]
\caption{\textbf{Runtime in GEUVADIS.} Mean wall-clock time per gene window in seconds for fixed-$K=5$ and BIC-stopped analyses.}
\label{tab:timing}
\centering\small
\begin{tabular}{lcc}
\toprule
Method & Fixed-$K$ & Adaptive (BIC) \\
\midrule
NaW & 0.47 & 3.38 \\
AsW & 0.56 & 3.60 \\
ExW & 13.40 & 42.49 \\
AsS & 0.57 & 3.69 \\
NaM & 7.54 & 3.70 \\
ExM & 21.84 & 6.81 \\
SuSiE & 2.87 & NA \\
LASSO & 3.33 & NA \\
\bottomrule
\end{tabular}
\end{table}

\begin{figure}[htbp]
\centering
\includegraphics[width=0.82\textwidth]{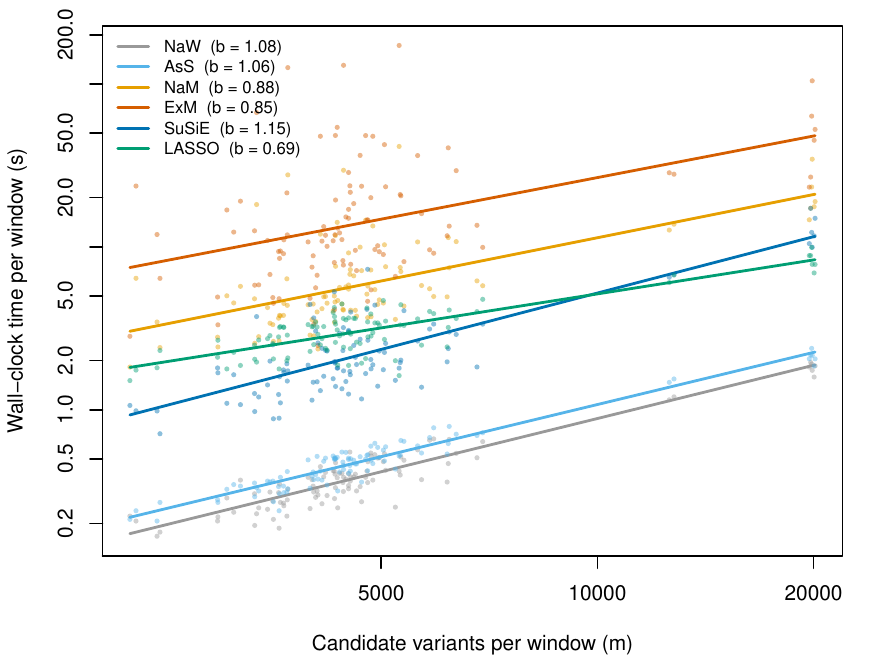}
\caption{\textbf{Empirical runtime scaling.} Per-window fixed-$K$ runtime against the number of candidate variants in GEUVADIS.}
\label{fig:scaling_m}
\end{figure}

Figure~\ref{fig:scaling_m} relates the fixed-$K$ runtime to the window dimension across the hundred GEUVADIS windows.

\subsection*{GEUVADIS cis-eQTL application and complementary uncertainty, selection-structure and sensitivity analyses}\label{subsec:supp_geuvadis}

\paragraph{Uncertainty in the primary support metric.}
Table~\ref{tab:geuvadis_ci} reports block-bootstrap uncertainty for the fixed-$K=5$ support rates, resampling $66$ disjoint genomic regions obtained by merging overlapping windows.

\begin{table}[htbp]
\caption{\textbf{Block-bootstrap confidence intervals for the LD-aware replication rates.} Mean LD-aware replication rate at fixed $K=5$ with $95\%$ percentile intervals. Last column: per-gene difference between the TSS-proximity and uniform priors. The BIC-stopping analogue gives the same picture.}
\label{tab:geuvadis_ci}
\centering\scriptsize
\setlength{\tabcolsep}{3pt}
\begin{tabular}{lccccc}
\toprule
\textbf{Method} & Uniform & PolyFun & TSS & Discordant & TSS $-$ Uniform \\
\midrule
NaW & 0.33 [0.28, 0.39] & 0.33 [0.27, 0.39] & 0.40 [0.34, 0.46] & 0.32 [0.26, 0.38] & 0.06 [0.04, 0.09] \\
AsW & 0.33 [0.28, 0.39] & 0.33 [0.27, 0.40] & 0.40 [0.33, 0.46] & 0.32 [0.26, 0.38] & 0.06 [0.04, 0.09] \\
ExW & 0.31 [0.26, 0.36] & 0.31 [0.26, 0.36] & 0.35 [0.30, 0.41] & 0.30 [0.25, 0.35] & 0.05 [0.03, 0.07] \\
AsS & 0.33 [0.28, 0.39] & 0.34 [0.28, 0.40] & 0.41 [0.35, 0.48] & 0.32 [0.27, 0.39] & 0.08 [0.05, 0.10] \\
NaM & 0.78 [0.70, 0.85] & 0.72 [0.63, 0.80] & 0.74 [0.66, 0.81] & 0.56 [0.46, 0.66] & -0.04 [-0.09, 0.00] \\
ExM & 0.78 [0.70, 0.84] & 0.72 [0.63, 0.80] & 0.73 [0.64, 0.80] & 0.57 [0.47, 0.66] & -0.04 [-0.09, -0.00] \\
SuSiE & 0.71 [0.63, 0.78] & 0.70 [0.63, 0.77] & 0.71 [0.63, 0.79] & 0.71 [0.63, 0.77] & 0.01 [-0.02, 0.03] \\
\bottomrule
\end{tabular}
\end{table}

\begin{figure}[htbp]
\centering
\includegraphics[width=0.82\textwidth]{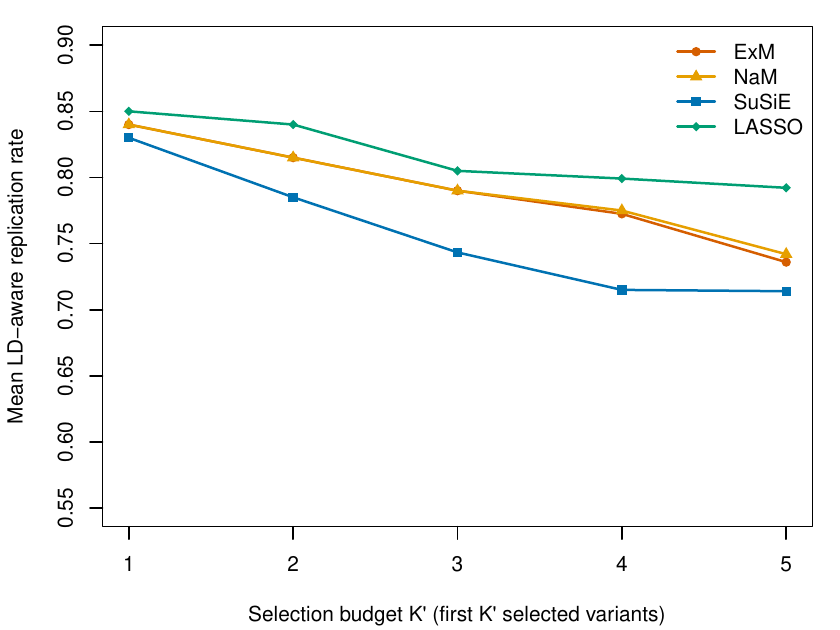}
\caption{\textbf{Cross-population support at matched selection budgets.} Mean LD-aware support rate of the first $K'$ selected variants for $K'=1,\ldots,5$ over the $100$ GEUVADIS windows under the TSS-proximity prior. LASSO is prior-free and is shown for its $K'$ strongest coefficients.}
\label{fig:budget_curve}
\end{figure}

\paragraph{Matched-budget comparison and redundancy.}
Figure~\ref{fig:budget_curve} compares the rankings at matched selection budgets, and Table~\ref{tab:geuvadis_clusters} reports the redundancy diagnostics behind the per-variant rates.

\begin{table}[htbp]
\caption{\textbf{Redundancy diagnostics of the fixed-$K=5$ selections.}}
\label{tab:geuvadis_clusters}
\centering\scriptsize
\setlength{\tabcolsep}{3pt}
\begin{tabular}{lcccc}
\toprule
Method & Uniform & PolyFun & TSS & Discordant \\
\midrule
NaW & 4.8 / 0.13 / 0.33 & 4.7 / 0.14 / 0.32 & 4.5 / 0.17 / 0.36 & 4.8 / 0.12 / 0.32 \\
AsW & 4.8 / 0.13 / 0.33 & 4.7 / 0.14 / 0.32 & 4.5 / 0.17 / 0.36 & 4.8 / 0.12 / 0.32 \\
ExW & 5.0 / 0.09 / 0.31 & 5.0 / 0.10 / 0.31 & 4.8 / 0.12 / 0.34 & 5.0 / 0.09 / 0.30 \\
AsS & 4.8 / 0.13 / 0.33 & 4.7 / 0.14 / 0.33 & 4.5 / 0.18 / 0.37 & 4.8 / 0.12 / 0.32 \\
NaM & 1.5 / 0.83 / 0.83 & 1.6 / 0.78 / 0.76 & 1.8 / 0.77 / 0.79 & 2.0 / 0.68 / 0.65 \\
ExM & 1.5 / 0.83 / 0.82 & 1.6 / 0.78 / 0.76 & 1.8 / 0.77 / 0.77 & 2.1 / 0.67 / 0.64 \\
SuSiE & 1.8 / 0.73 / 0.76 & 1.8 / 0.73 / 0.75 & 1.8 / 0.73 / 0.77 & 1.9 / 0.72 / 0.76 \\
\bottomrule
\end{tabular}
\end{table}

\paragraph{Prediction and conditional replication.}
Table~\ref{tab:geuvadis_r2} reports the out-of-sample adjusted $R^2$ in YRI, and Table~\ref{tab:geuvadis_cond} applies the stricter conditional-replication definition, testing each selected variant given the variants selected before it.

\begin{table}[htbp]
\caption{\textbf{Out-of-sample adjusted $R^2$ in the YRI replication panel.} Mean adjusted $R^2$ of the selected set over the $100$ genes, by prior specification, at fixed $K=5$ (top block) and under BIC stopping (bottom block, mean selected size $\hat K$ in parentheses). The larger single-SNP selections under BIC buy a larger out-of-sample $R^2$ at a lower per-variant replication rate (Table~\ref{tab:geuvadis_adaptive}). The prior-free LASSO attains $0.270$ with $\hat K=3.8$.}
\label{tab:geuvadis_r2}
\centering\small
\begin{tabular}{lcccc}
\toprule
Method & Uniform & PolyFun & TSS & Discordant \\
\midrule
\multicolumn{5}{l}{\emph{Fixed $K=5$}} \\
NaW & 0.243 & 0.242 & 0.255 & 0.238 \\
AsW & 0.243 & 0.242 & 0.255 & 0.238 \\
ExW & 0.233 & 0.233 & 0.243 & 0.230 \\
AsS & 0.241 & 0.242 & 0.256 & 0.239 \\
NaM & 0.266 & 0.259 & 0.261 & 0.238 \\
ExM & 0.266 & 0.257 & 0.260 & 0.240 \\
SuSiE & 0.253 & 0.250 & 0.254 & 0.251 \\
\midrule
\multicolumn{5}{l}{\emph{BIC stopping}} \\
NaW & 0.253 (15.6) & 0.249 (10.7) & 0.254 (8.3) & 0.250 (14.6) \\
AsW & 0.256 (15.5) & 0.247 (10.6) & 0.254 (8.5) & 0.251 (14.4) \\
ExW & 0.244 (13.4) & 0.242 (9.3) & 0.247 (8.7) & 0.237 (13.2) \\
AsS & 0.256 (15.5) & 0.249 (10.5) & 0.255 (8.2) & 0.253 (14.3) \\
NaM & 0.236 (1.8) & 0.226 (1.5) & 0.234 (1.4) & 0.221 (1.5) \\
ExM & 0.237 (1.8) & 0.227 (1.5) & 0.234 (1.4) & 0.220 (1.5) \\
\bottomrule
\end{tabular}
\end{table}

\begin{table}[htbp]
\caption{\textbf{Conditional replication sensitivity (fixed $K=5$).}}
\label{tab:geuvadis_cond}
\centering\small
\begin{tabular}{lcccc}
\toprule
Method & Uniform & PolyFun & TSS & Discordant \\
\midrule
NaW & 0.220 & 0.228 & 0.228 & 0.214 \\
AsW & 0.220 & 0.226 & 0.226 & 0.214 \\
ExW & 0.210 & 0.210 & 0.218 & 0.206 \\
AsS & 0.220 & 0.224 & 0.224 & 0.222 \\
NaM & 0.240 & 0.226 & 0.230 & 0.200 \\
ExM & 0.242 & 0.224 & 0.228 & 0.196 \\
SuSiE & 0.214 & 0.214 & 0.214 & 0.206 \\
\bottomrule
\end{tabular}
\end{table}

\paragraph{Sensitivity to prior amplitude and replication definition.}
Table~\ref{tab:geuvadis_prior_scale} varies the amplitude of the prior weights, and Table~\ref{tab:geuvadis_sensitivity} reports the exact-variant and per-window Bonferroni readings of the fixed-$K$ analysis.

\begin{table}[htbp]
\caption{\textbf{Sensitivity to the amplitude of the prior weights.} Mean LD-aware replication rate of the fixed-$K=5$ selections (nominal test) when the three-level prior scale of the main analysis, $\{0.3, 0.5, 0.7\}$, is replaced by a weaker $\{0.4, 0.5, 0.6\}$ or a stronger $\{0.2, 0.5, 0.8\}$ contrast, level by level. The uniform arm is constant at $0.5$ and hence invariant by construction. Its repetition across scales is an internal consistency check.}
\label{tab:geuvadis_prior_scale}
\centering\small
\setlength{\tabcolsep}{4pt}
\begin{tabular}{llcccc}
\toprule
Method & Scale & Uniform & PolyFun & TSS & Discordant \\
\midrule
\multirow{3}{*}{AsS} & $\{0.4, 0.5, 0.6\}$ & 0.332 & 0.340 & 0.356 & 0.330 \\
& $\{0.3, 0.5, 0.7\}$ & 0.332 & 0.338 & 0.408 & 0.324 \\
& $\{0.2, 0.5, 0.8\}$ & 0.332 & 0.336 & 0.500 & 0.314 \\
\addlinespace
\multirow{3}{*}{ExW} & $\{0.4, 0.5, 0.6\}$ & 0.308 & 0.314 & 0.310 & 0.298 \\
& $\{0.3, 0.5, 0.7\}$ & 0.308 & 0.312 & 0.354 & 0.298 \\
& $\{0.2, 0.5, 0.8\}$ & 0.308 & 0.300 & 0.422 & 0.290 \\
\addlinespace
\multirow{3}{*}{NaM} & $\{0.4, 0.5, 0.6\}$ & 0.780 & 0.728 & 0.754 & 0.604 \\
& $\{0.3, 0.5, 0.7\}$ & 0.780 & 0.720 & 0.740 & 0.562 \\
& $\{0.2, 0.5, 0.8\}$ & 0.780 & 0.720 & 0.740 & 0.554 \\
\addlinespace
\multirow{3}{*}{ExM} & $\{0.4, 0.5, 0.6\}$ & 0.778 & 0.726 & 0.752 & 0.596 \\
& $\{0.3, 0.5, 0.7\}$ & 0.778 & 0.718 & 0.734 & 0.566 \\
& $\{0.2, 0.5, 0.8\}$ & 0.778 & 0.718 & 0.730 & 0.548 \\
\bottomrule
\end{tabular}
\end{table}

\begin{table}[htbp]
\caption{\textbf{Replication sensitivities of the fixed-$K=5$ selections.} Each cell reports, by prior specification, the exact-variant replication rate (crediting a selected variant only when it replicates itself, no LD credit), followed by the LD-aware rate under a per-window Bonferroni replication threshold $\alpha/m$, at which the $n=89$ replication set is underpowered. Both variants preserve the ordering of priors and methods of Table~\ref{tab:geuvadis_fixedK} of the main text.}
\label{tab:geuvadis_sensitivity}
\centering\small
\setlength{\tabcolsep}{4pt}
\begin{tabular}{lcccc}
\toprule
Method & Uniform & PolyFun & TSS & Discordant \\
\midrule
NaW & 0.270 / 0.162 & 0.276 / 0.166 & 0.322 / 0.190 & 0.258 / 0.164 \\
AsW & 0.270 / 0.162 & 0.274 / 0.166 & 0.320 / 0.190 & 0.258 / 0.164 \\
ExW & 0.238 / 0.142 & 0.250 / 0.138 & 0.280 / 0.160 & 0.226 / 0.140 \\
AsS & 0.270 / 0.160 & 0.276 / 0.168 & 0.328 / 0.196 & 0.264 / 0.164 \\
NaM & 0.728 / 0.496 & 0.652 / 0.462 & 0.666 / 0.462 & 0.472 / 0.314 \\
ExM & 0.728 / 0.498 & 0.652 / 0.462 & 0.660 / 0.462 & 0.480 / 0.308 \\
SuSiE & 0.632 / 0.414 & 0.616 / 0.402 & 0.630 / 0.414 & 0.618 / 0.408 \\
\bottomrule
\end{tabular}
\end{table}

\end{document}